\documentclass[%
reprint,
superscriptaddress,
showpacs,preprintnumbers,
amsmath,amssymb,
aps,
pra,
longbibliography
]{revtex4-2}
\usepackage{color}
\usepackage{graphicx}
\usepackage{bm}
\usepackage{mathrsfs}
\usepackage{dsfont}
\usepackage{ulem}
\usepackage[colorlinks,allcolors=blue]{hyperref}
\usepackage{subfigure}
\usepackage{float}

\makeatletter

\newcommand{\Rmnum}[1]{\expandafter\@slowromancap\romannumeral #1@}

\makeatother

\begin{document}
%\preprint{}

\title{ Multiparameter simultaneous optimal estimation with an $SU(2)$ coding unitary evolution}


\author{Yu Yang}
\email{yangyu1229@hotmail.com}
\affiliation{Ministry of Education Key Laboratory for Nonequilibrium Synthesis 
	and Modulation of Condensed Matter, Shaanxi Province Key Laboratory 
	of Quantum Information and Quantum Optoelectronic Devices, School of 
	Physics, Xi’an Jiaotong University, Xi’an 710049, China}
\author{Shihao Ru}
\affiliation{Ministry of Education Key Laboratory for Nonequilibrium Synthesis 
	and Modulation of Condensed Matter, Shaanxi Province Key Laboratory 
	of Quantum Information and Quantum Optoelectronic Devices, School of 
	Physics, Xi’an Jiaotong University, Xi’an 710049, China}
\author{Min An}
\affiliation{Ministry of Education Key Laboratory for Nonequilibrium Synthesis 
	and Modulation of Condensed Matter, Shaanxi Province Key Laboratory 
	of Quantum Information and Quantum Optoelectronic Devices, School of 
	Physics, Xi’an Jiaotong University, Xi’an 710049, China}
\author{Yunlong Wang}
\affiliation{Ministry of Education Key Laboratory for Nonequilibrium Synthesis 
	and Modulation of Condensed Matter, Shaanxi Province Key Laboratory 
	of Quantum Information and Quantum Optoelectronic Devices, School of 
	Physics, Xi’an Jiaotong University, Xi’an 710049, China}
\author{Feiran Wang}
\affiliation{Ministry of Education Key Laboratory for Nonequilibrium Synthesis 
	and Modulation of Condensed Matter, Shaanxi Province Key Laboratory 
	of Quantum Information and Quantum Optoelectronic Devices, School of 
	Physics, Xi’an Jiaotong University, Xi’an 710049, China}
\affiliation{School of Science, Xi’an Polytechnic University, Xi’an 710048, China}
\author{Pei Zhang}
\affiliation{Ministry of Education Key Laboratory for Nonequilibrium Synthesis 
	and Modulation of Condensed Matter, Shaanxi Province Key Laboratory 
	of Quantum Information and Quantum Optoelectronic Devices, School of 
	Physics, Xi’an Jiaotong University, Xi’an 710049, China}
\author{Fuli Li}
\email{flli@xjtu.edu.cn}
\affiliation{Ministry of Education Key Laboratory for Nonequilibrium Synthesis 
	and Modulation of Condensed Matter, Shaanxi Province Key Laboratory 
	of Quantum Information and Quantum Optoelectronic Devices, School of 
	Physics, Xi’an Jiaotong University, Xi’an 710049, China}

%\date{\today}

\begin{abstract}
In a ubiquitous $SU(2)$ dynamics, achieving the simultaneous optimal estimation of multiple parameters is significant but difficult. 
Using quantum control to optimize this $SU(2)$ coding unitary evolution is one of solutions.
We propose a method, characterized by the nested cross-products of the coefficient vector $\mathbf{X}$ of $SU(2)$ generators and its partial derivative $\partial_\ell \mathbf{X}$, to investigate the control-enhanced quantum multiparameter estimation.
 Our work reveals that quantum control is not always functional in improving the estimation precision, which depends on the characterization of an $SU(2)$ dynamics with respect to the objective parameter.  
This characterization is quantified by  the angle $\alpha_\ell$ between $\mathbf{X}$ and $\partial_\ell \mathbf{X}$.
For an $SU(2)$ dynamics featured by $\alpha_\ell=\pi/2$, the promotion of the estimation precision can get the most benefits from the controls. 
When $\alpha_\ell$ gradually closes to $0$ or $\pi$, the precision promotion contributed to by quantum control correspondingly becomes inconspicuous. 
Until a dynamics with $\alpha_\ell=0$ or $\pi$, quantum control completely loses its advantage. 
In addition, we find a set of conditions restricting the simultaneous optimal estimation of all the parameters, but fortunately,  which can be removed by using a maximally entangled two-qubit state as the probe state and adding an ancillary channel into the configuration.
Lastly, a spin-$1/2$ system is taken as an example to verify the above-mentioned conclusions.
Our proposal sufficiently exhibits the hallmark of control-enhancement in fulfilling the multiparameter estimation mission, and it is applicable to an arbitrary $SU(2)$ parametrization process.
\end{abstract}

%\keywords{}

\maketitle

\section{Introduction}
Much attention to multiparameter estimation missions by virtue of quantum resource~\cite{Multi_1,Multi_2,Multi_3} have been paid  in recent years, such as studying the magnetometry~\cite{Magneticfile0}, developing the gyroscope~\cite{PhysRevA.95.012326,PhysRevApplied.14.034065} and designing the quantum network~\cite{PhysRevLett.120.080501}. 
Some specific issues like estimating relevant parameters in the quantum interferometer~\cite{Pa,PhysRevA.interferometer}, recovering the position information of two incoherent point sources~\cite{PhysRevX.6.031033,PhysRevLett.122.140505}, achieving the ultimate timing resolution~\cite{PRXQuantum.2.010301}, or estimating the temperature and pressure by the nitrogen-vacancy (NV) center in diamond~\cite{PhysRevLett.112.047601},  were studied.
Thereinto, a widely discussed example is estimating the attributes of an unknown magnetic field in different physical ensembles~\cite{Pang2014, Magneticfile1}. 
However, we notice that these seemingly different dynamics can be summarized as a class of unitary evolution expressed by the $SU(2)$ group, the corresponding generator (a time-independent Hamiltonian) belongs to $SU(2)$ algebra. 
Ref.~\cite{PhysRevA.92.012312} proposed a representation for the $SU(2)$ single-parameter estimation problem and clarified that multiple parameters cannot be simultaneously estimated.

In quantum multiparameter estimation, the probe state $\hat{\rho}_\text{in}$ evolves into $\hat{\rho}_\mathbf{x}$  under a parameter-dependent dynamics that carries a set of to-be-estimated parameters 
$\mathbf{x}=\{x_1,x_2,\cdots x_d\}$. 
The encoded state $\hat{\rho}_\mathbf{x}$ is measured by a positive-operator-valued measurement (POVM) $\{\hat{\Pi}_z|\hat{\Pi}_z \ge 0, \sum_z \hat{\Pi}_z=\hat{I}\}$, where $\hat{I}$ is the identity operator.
Then one obtains the mesurement result $z$ with a probability $p(z|\mathbf{x})=\text{Tr}[\hat{\Pi}_z \hat{\rho}_\mathbf{x}]$ according to the Born rule.
Employing a local unbaised estimator to deal with $z$, the estimation values of $\mathbf{x}$ can be extrapolated.
The ultimate estimation precision of each parameter is bounded below by the reciprocal of its quantum Fisher information (QFI). 
The whole estimation precision with respect to all the parameters is bounded below by the inverse of quantum Fisher information matrix (QFIM).
The above dynamics is usually constructed with a parallel structure~\cite{Multi_2,PhysRevLett.125.020501} or a sequential structure~\cite{Se,Multi_Sequential}. 
However,  there is a roadblock hindering the realistic application of the theory of quantum multiparameter estimation. 
That is the simultaneous optimal estimation of multiple parameters generally cannot be achieved even asymptotically~\cite{Multi_1,Multi_3}. 
It inevitably gives rise to a trade-off among multiple estimation precision.
To circumvent this difficulty, pursing the more compatible probe state and the measurement scheme are two feasible strategies~\cite{Mu_satu3,Belliardo_2021,Kull_2020,albarelli2021probe}.  
Apart from that, employing quantum control to optimize the original dynamics so as to eliminate this trade-off, is another promising solution.
Ref.~\cite{Se}  concentrated on a specific problem of estimating the magnetic field, and presented that introducing a set of controls into the sequential scheme can achieve the above aim.
The subsequent experiments~\cite{PhysRevLett.125.020501,Multi_Sequential}  demonstrated this proposal.
In Ref.~\cite{Se}, the proposed analytical method bridges the maximal QFIM and the difference between the maximal and minimal eigenangles of a unitary operator.
In addition, Refs.~\cite{Pang2014,pang2017optimal}  also accounted for the function of quantum control  in the single-parameter estimation missions.
However, we notice that there is not a relatively complete work to discuss the role of quantum control in a universal $SU(2)$ coding unitary evolution, especially to study under which kind of circumstance quantum control can play full advantage or lose its ability.

Differently from the previous works~\cite{Pang2014,Se}, we present  a method based on the nested cross-products of the coefficient vector $\mathbf{X}$ of $SU(2)$ generators and its partial derivative $\partial_\ell \mathbf{X}$, to investigate control-enhanced quantum multiparameter estimation. 
It  is applicable for an arbitrary $SU(2)$ parametrization process.
Our work reveals that quantum control is not always functional in improving the estimation precision, which dominatly depends on the characterization of an $SU(2)$ dynamics with respect to the objective parameter.  
This characterization is quantified by the angle $\alpha_\ell$ between $\mathbf{X}$ and $\partial_\ell \mathbf{X}$.
Concretely, for the parameter $x_\ell$, if an $SU(2)$ dynamics is featured by $\alpha_\ell=\pi/2$, the promotion of the estimation precision can get the most benefits from the controls. 
When $\alpha_\ell$ closes to $0$ or $\pi$, the power of quantum control will correspondingly decreases.
Until for an $SU(2)$ dynamics with $\alpha_\ell=0$ or $\pi$, quantum control completely loses its advantage. 
Furthermore, the attainability of simultaneous optimal estimation in an $SU(2)$ dynamics  is rather pivotal. 
By analyzing the QFI maximum and the weak commutation condition, 
we therewith find a set of conditions that restrict the expected simultaneous optimal estimation.
But fortunately using a maximally entangled two-qubit state as the probe state and adding an ancillary channel into the setup can remove these confines.

The remainder of this paper is organized as follows. 
Section~\ref{Sec:QETR} gives a brief review about the theory of quantum multiparameter estimation.  
Then our analytical method is proposed in Sec.~\ref{Sec:Scheme}, based on this we present a set of conditions restricting the simultaneous optimal estimation in the no-control scheme (see Sec.~\ref{Sec:withoutC}) and in the control-enhanced scheme (see Sec.~\ref{Sec:withC}).  
In Sec.~\ref{Sec:withC}, specifically, the effectiveness of quantum control for different values of $\alpha_\ell$ are described in detail.
After that the attainability of the control-enhanced simultaneous optimal estimation is discussed at Sec.~\ref{Sec:Ancillary}.
In Sec.~\ref{Sec:Example}, we take a spin-$1/2$ system as an example to validate the correctness of our proposal.
It is followed by the conclusion of this paper in Sec.~\ref{Sec:Conclusion}.

\section{Quantum multiparameter estimation theory}\label{Sec:QETR}
Most of the realistic scenarios need to estimate multiple parameters as precise as possible.
Unknown parameters are $\mathbf{x}=\{x_1,x_2,\cdots x_d\} \in \Theta$ that is an open subset of $\mathbb{R}^d$.
The ultimate estimation precision is bounded below by the quantum Cram{$\acute{e}$}r Rao Bound (QCRB) as
\begin{eqnarray}\label{Multi_QCRB}
	\text{Cov}\left(\tilde{\mathbf{x}} \right) \ge \frac{1}{M} {\mathcal{F}}^{-1}\;,
\end{eqnarray}
where $\text{Cov}\left(\tilde{\mathbf{x}} \right)$ is the error covariance matrix of the local unbiased estimator $\tilde{\mathbf{x}}=\{\tilde{x}_1,\tilde{x}_2,\cdots \tilde{x}_d\}$, $\left(\text{Cov}(\tilde{\mathbf{x}})\right)_{\ell \ell'}=\text{E}[\tilde{x}_\ell \tilde{x}_{\ell'}]-\text{E}[\tilde{x}_\ell] \text{E}[\tilde{x}_{\ell'}]$ ($\text{E}[\bullet]$ denotes the expectation).
$M$ represents the number of times that the estimation procedure is repeated, and $\mathcal{F}^{-1}$ is the inverse of the QFIM $\mathcal{F}$. Eq.~(\ref{Multi_QCRB}) states that $\text{Cov}\left(\tilde{\mathbf{x}} \right) - \mathcal{F}^{-1}/M$ is a positive semidefinite matrix.
The precision limit of the parameter $x_\ell$ reads
\begin{eqnarray}\label{tmp1}
	\Delta^2 x_\ell \ge \frac{1}{M} \left(\mathcal{F}^{-1}\right)_{\ell\ell} \ge \frac{1}{M \mathcal{F}_{\ell\ell}}\;,
\end{eqnarray}
where $\Delta x_\ell$ is the standard deviation of $x_\ell$.
 $\left(\mathcal{F}^{-1}\right)_{\ell\ell}  \ge 1/\mathcal{F}_{\ell\ell}$ since
$\mathcal{F}$ is positive semidefinite, and the equality holds only for a diagonal $\mathcal{F}$~\cite{Review}.

Now we introduce a nontrivial Hermitian operator $\hat{\mathcal{H}}_\ell$ with respect to $x_\ell$ \cite{liu2015quantum,Review}
\begin{eqnarray}\label{eq:def}
	\hat{\mathcal{H}}_\ell:=i(\partial_\ell \hat{U}^\dagger)\hat{U}\;,
\end{eqnarray}
where $\hat{U}$ is the unitary transformation from the probe state $\hat{\rho}_\text{in}$ to the encoded state $\hat{\rho}_\mathbf{x}$ and $\partial_\ell \hat{U}^\dagger:=\partial \hat{U}^\dagger/\partial x_\ell$.
Taking a two-dimensional pure state as $\hat{\rho}_{\text{in}}$, one can write the $(\ell,\ell')$-th entry of the QFIM $\mathcal{F}$ as~\cite{liu2015quantum}
\begin{eqnarray}\label{eq:cov_pure}
	\mathcal{F}_{\ell \ell'}=4\text{Cov}_{\text{in}}(\hat{\mathcal{H}}_\ell,\hat{\mathcal{H}}_{\ell'})\;,
\end{eqnarray}
with
\begin{eqnarray}\label{eq:cov}
\text{Cov}_\text{in} (\hat{\mathcal{H}}_\ell,\hat{\mathcal{H}}_{\ell'})  \!=\!  \frac{1}{2}\text{Tr} \left[\{\hat{\mathcal{H}}_\ell, \hat{\mathcal{H}}_{\ell'} \} \hat{\rho}_\text{in} \right] \!-\! \text{Tr}\! \left[\hat{\mathcal{H}}_\ell \hat{\rho}_\text{in}  \hat{\mathcal{H}}_{\ell'} \hat{\rho}_\text{in}  \right],\nonumber\\
\end{eqnarray}
where $\{\bullet,\bullet\}$ denotes the anticommutator. 
From Eq.~(\ref{eq:cov_pure}), the $\ell$-th diagonal element of the QFIM $\mathcal{F}$ is
\begin{eqnarray}\label{eq:QFIM_pure}
	\mathcal{F}_{\ell\ell}=4\text{Tr}\left[\Delta^2 \hat{\mathcal{H}_\ell}\; \hat{\rho}_{\text{in}} \right]\;,
\end{eqnarray}
with 
\begin{eqnarray}\label{eq:var}
	\text{Tr}\left[\Delta ^2 \hat{\mathcal{H}}_\ell \;  \hat{\rho}_\text{in} \right]=\text{Tr}\left[ \hat{\mathcal{H}}_{\ell}^2 \; \hat{\rho}_\text{in} \right]-\text{Tr}^2 \left[\hat{\mathcal{H}}_\ell \; \hat{\rho}_\text{in} \right]\;.
\end{eqnarray}

Apart from improving the estimation precision of each parameter so as to approach the QCRB as much as possible, 
whether these QCRBs can be simultaneously reached (i.e., whether multiple parameters can be optimally estimated at the same time) is also a significant problem.
In quantum single parameter estimation, the QCRB can be asymptotically reached~\cite{QFI}, while for quantum multiparameter estimation it is not always attainable unless the weak commutation condition~\cite{Mu_satu,Mu_satu2,Mu_satu3} 
\begin{eqnarray}\label{eq:weak_comm}
	\text{Tr}\left[\left[\hat{L}_\ell,\hat{L}_{\ell'}\right]\hat{\rho}_{\mathbf{x}}\right]=0 \quad \forall \ell\neq {\ell'} \;,
\end{eqnarray}
is satisfied. Here $\ell$ and $\ell'$ traverse the indexes of all the parameters. $\hat{L}_\ell$ is a symmetric logarithmic derivative (SLD) operator with respect to $x_\ell$, which is defined by $\partial_\ell \hat{\rho}_\mathbf{x}=(\hat{\rho}_\mathbf{x} \hat{L}_\ell +\hat{L}_\ell \hat{\rho}_\mathbf{x} )/2$ with a shorthand notation $\partial_\ell \hat{\rho}_\mathbf{x}:=\partial \hat{\rho}_\mathbf{x}/\partial x_\ell$.
Eq.~(\ref{eq:weak_comm}) is applicable to a pure~\cite{Mu_satu} or a mixed~\cite{Mu_satu3} encoded state.
According to Eq.~(\ref{eq:def}), Eq.~(\ref{eq:weak_comm}) can be renewed as 
(see Appendix~\ref{Sec:AppA})
\begin{eqnarray}\label{eq:wcc}
	\text{Tr}\left[ \left[\hat{\mathcal{H}}_\ell,\hat{\mathcal{H}}_{\ell'} \right] \hat{\rho}_\text{in}\right]=0 \quad \forall \ell \neq {\ell'} \;,
\end{eqnarray}
which implies a sufficient but not necessary condition $ \left[\hat{\mathcal{H}}_\ell,\hat{\mathcal{H}}_{\ell'} \right]=0$.

\section{Simultaneous optimal estimation with an $SU(2)$ coding unitary evolution}\label{Sec:Scheme}
\begin{figure}[!h]
	\centering
	\includegraphics[width=0.47\textwidth]{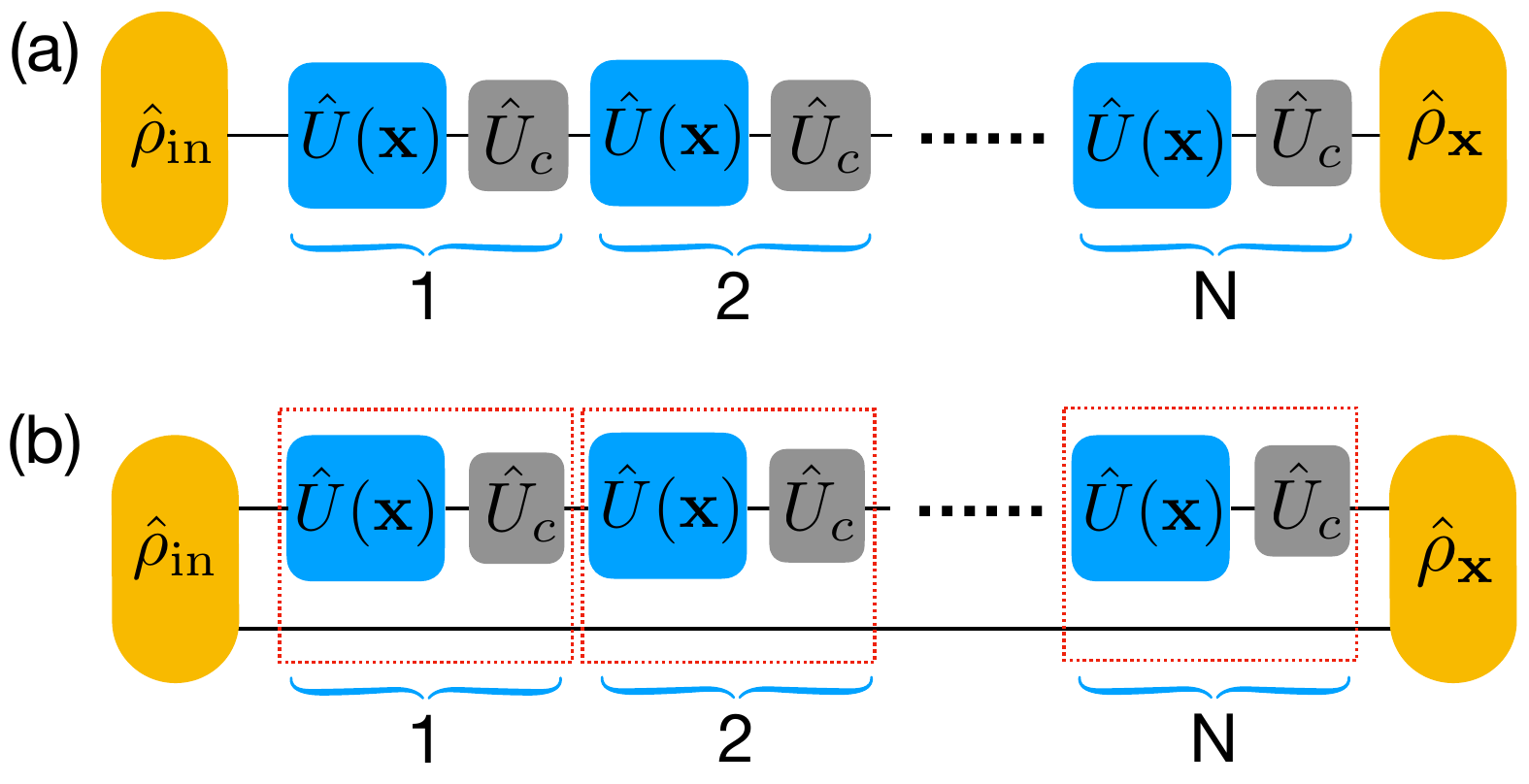}
	\caption
	{The control-enhanced sequential scheme for quantum multiparameter estimation (a) without the ancillary channel and (b) with an ancillary channel. 	
	The following measurement procedure and the data processing do not be shown.
	$\hat{\rho}_{\text{in}}$ and $\hat{\rho}_\mathbf{x}$ separately denote the probe state and the encoded state.
	$\mathbf{x}=\{x_1,x_2, x_3\}$ is a set of to-be-estimated parameters.
	The whole dynamics evolution is divided into $N$ groups that each one includes an $SU(2)$ parametrization process $\hat{U}(\mathbf{x})$  and a control operation $\hat{U}_c$.}
	\label{Fig_scheme}
\end{figure}

A sequential scheme of estimating multiple parameters is depicted in Fig.~\ref{Fig_scheme} (a).  
The initial probe state $\hat{\rho}_{\text{in}}$  evolves through  $N$ groups dynamics successively. 
The dynamics of each group includes an $SU(2)$ parametrization process $\hat{U}(\mathbf{x})=e^{-it\hat{H}(\mathbf{x})}$  and a control operation $\hat{U}_c=e^{-it\hat{H}_c}$. 
The action time $t$  makes up a whole evolution time as $T=N t$.
Setting $\hat{U}_c=\hat{I}$ can be used to mimic the scheme without controls, and setting $\hat{U}_c$ to be some special forms such as $\hat{U}_c=\hat{U}^\dagger(\mathbf{x})$~\cite{Se, Single_Sequential_Hou,Multi_Sequential},  can be used to investigate the control-enhanced quantum multiparameter estimation.
$N$ groups unitary transformations cascading with each other are employed to achieve the simultaneous optimal estimation of $\mathbf{x}$ and to arrive the Heisenberg scaling $1/N$ or $1/T$~\cite{giovannetti2004quantum,QM} of each parameter.
In Fig.~\ref{Fig_scheme} (a), the entire unitary transformation from $\hat{\rho}_\text{in}$ to $\hat{\rho}_\mathbf{x}$ can be expressed by
\begin{eqnarray}\label{eq:total_evolution}
	\hat{U}_{tot}=\left(\hat{U}_c \hat{U}(\mathbf{x})\right)^N
	 \simeq \left(e^{-it(\hat{H}(\mathbf{x})+\hat{H}_c)}\right)^N\;,
\end{eqnarray}
where in the last approximation, one can omit the high-order term with respect to $t$ when $t$ is sufficiently small.

Now we consider a general time-independent Hamiltonian given by the linear combination of $SU(2)$ generators, 
\begin{eqnarray}\label{eq:HSU(2)}
\hat{H}(\mathbf{x})=\mathbf{X} \cdot \vec{J}\;,
\end{eqnarray} 
where 
$\mathbf{X}=(X_1(\mathbf{x}),X_2(\mathbf{x}),X_3(\mathbf{x}))$ is a three-dimensional vector and $X_\varepsilon(\mathbf{x})$ represents a function of $\mathbf{x}$ with $\varepsilon=1,2,3$. 
$\vec{J}=(\hat{j}_1,\hat{j}_2,\hat{j}_3)$ are three generators of $SU(2)$ algebra (i.e., $\mathfrak{su}(2)$), which obey the commutation relation
\begin{eqnarray}
\left[\hat{j}_m, \hat{j}_k \right]=i \epsilon_{mkl}\; \hat{j}_l\;,
\end{eqnarray}
with the Levi-Civita symbol $\epsilon_{mnl}$ and 
\begin{eqnarray}
\left(\vec{a} \cdot \vec{J}\right)   \left( \vec{b} \cdot \vec{J} \right)=\frac{1}{4}(\vec{a} \cdot \vec{b}) \hat{I} +\frac{i}{2} (\vec{a} \times \vec{b}) \cdot \vec{J}\label{eq:pro}\;,
\end{eqnarray}
where $\vec{a}$ and $\vec{b}$ are two arbitrary three-dimensional vectors.
For an $SU(2)$ dynamics, the number of to-be-estimated parameters is restricted to three.

We then consider the Hamiltonian of quantum control
\begin{eqnarray}\label{eq:Hc}
	\hat{H}_c=\mathbf{X}_c \cdot \vec{J}\;,
\end{eqnarray}
where $\mathbf{X}_c=(X_1^{(c)}(\mathbf{x}_c),X_2^{(c)}(\mathbf{x}_c),X_3^{(c)}(\mathbf{x}_c))$ is a three-dimensional vector and $X_\varepsilon^{(c)}(\mathbf{x}_c)$ denotes a function of $\mathbf{x}_c$ with $\varepsilon=1,2,3$. 
$\mathbf{x}_c=\{\tilde{x}_1,\tilde{x}_2, \tilde{x}_3\}$ is a set of estimated values of $\mathbf{x}$ used in the control. 
Accordingly, in Fig.~\ref{Fig_scheme} (a), the Hamiltonian of the $k$-th unitary cell is
\begin{eqnarray}\label{eq:Hk}
\hat{H}_k=\hat{H}(\mathbf{x})+\hat{H}_c=\mathbf{S}\cdot \vec{J}\;,
\end{eqnarray}
with $\mathbf{S}=\mathbf{X}+\mathbf{X}_c$.
To achieve the simultaneous optimal estimation of $\mathbf{x}$, two cases without and with control-enhancement are respectively discussed in the following.
\subsection{Without controls}\label{Sec:withoutC}
We assume $\hat{U}_c$ involved in Fig.~\ref{Fig_scheme} (a) to be an identity operator  to mimic the scheme without controls.
From Eqs.~(\ref{eq:def}) and (\ref{eq:total_evolution}), we have
\begin{eqnarray}\label{eq:partH1}
\hat{\mathcal{H}}_\ell&=&i(\partial_\ell \hat{U}_{tot}^\dagger)\hat{U}_{tot}\nonumber\\
&=&i(\partial_\ell e^{it \sum_{k=1}^{N} \hat{H}_k})e^{-it \sum_{k=1}^{N} \hat{H}_k}\nonumber\\
&=&i  \int_0^1 \! ds \; e^{ ist \sum_{k=1}^{N} \hat{H}_k} \partial_\ell  \! \left( it \sum_{k=1}^{N} \hat{H}_k \right) \!
e^{i(1-s) t \sum_{k=1}^{N} \hat{H}_k } \nonumber\\
&\times& e^{-it \sum_{k=1}^{N} \hat{H}_k }\nonumber\\
&=&-\Bigg\{ t\partial_\ell \sum_{k=1}^N \hat{H}_k +\frac{i t^2}{2} \left[ \sum_{k=1}^N \hat{H}_k,\partial_\ell  \sum_{k=1}^{N} \hat{H}_k\right]\nonumber\\
&-&\frac{t^3}{3!} \left[ \sum_{k=1}^N \hat{H}_k, \left[ \sum_{k=1}^N \hat{H}_k,\partial_\ell  \sum_{k=1}^{N} \hat{H}_k\right]  \right]+\cdots 
\Bigg\}\nonumber\\
&=& i\sum_{n=0}^{\infty} \frac{(it)^{n+1}}{(n+1)!} \left( \sum_{k=1}^{N} \hat{H}_k \right)^{\times n} \left(\partial_\ell  \sum_{k=1}^{N} \hat{H}_k \right)\;,
\end{eqnarray}
where the notation $\hat{A}^{\times n}(\bullet):=\overbrace{[\hat{A},[\hat{A},\cdots [\hat{A}}^{n \; \text{times}},\bullet]$ is introduced to express the nested commutators. In the above calculations, the derivative of an exponential operator $\partial_\ell e^{\hat{A}}=\int_0^1ds\; e^{s\hat{A}} \left(\partial_\ell \hat{A}\right)e^{(1-s)\hat{A}}$ and the well-known expansion $e^{\hat{A}} \hat{B} e^{-\hat{A}}=\hat{B}+[\hat{A},\hat{B}]+(1/2!)[\hat{A}, [\hat{A},\hat{B}]]+\cdots$~\cite{wilcox1967exponential} are employed.  
According to 
$\sum_{k=1}^{N} \hat{H}_k=N \hat{H}(\mathbf{x})$ and $\partial_\ell \sum_{k=1}^{N}\hat{H}_k=N\left(\partial_\ell \hat{H}({\mathbf{x}})\right)$, Eq.~(\ref{eq:partH1}) is rewritten as
\begin{eqnarray}\label{eq:Hwithout}
\hat{\mathcal{H}}_\ell=i\sum_{n=0}^{\infty} \frac{(iT)^{n+1}}{(n+1)!} \hat{H}(\mathbf{x})^{\times n} \left( \partial_\ell \hat{H}(\mathbf{x}) \right)\;.
\end{eqnarray}
According to Eq.~(\ref{eq:HSU(2)}), we also have\\
\begin{eqnarray}
\partial_\ell \hat{H}(\mathbf{x})&=&\partial_\ell \mathbf{X} \cdot \vec{J}\;,\label{eq:Multi1}\\
\left[\hat{H}(\mathbf{x}),\partial_\ell \hat{H}(\mathbf{x})\right]&=&i\left(\mathbf{X} \times \partial_\ell \mathbf{X}\right) \cdot \vec{J}\;,\\
\hat{H}(\mathbf{x})^{\times n} \left(\partial_\ell \hat{H}(\mathbf{x})\right)&=&i^n \left( \mathbf{X}_{\times n} \left( \partial_\ell \mathbf{X}\right) \right) \cdot \vec{J}\;,\label{eq:ntimes}
\end{eqnarray}
where the notation $\mathbf{Z}_{\times n} \left( \bullet \right):=\overbrace{(\mathbf{Z} \times (\mathbf{Z} \times \cdots (\mathbf{Z}}^{n\; \text{times}} \times  \bullet)$ is used to denote the nested cross-products. 
Substituting Eq.~(\ref{eq:ntimes}) into Eq.~(\ref{eq:Hwithout}), one can obtain
\begin{eqnarray}\label{eq:Hnew}
\hat{\mathcal{H}}_\ell=\sum_{n=0}^{\infty} \frac{(-T)^{n+1}}{(n+1)!}
\left( \mathbf{X}_{\times n} \left( \partial_\ell \mathbf{X}\right) \right) \cdot \vec{J}\;.
\end{eqnarray}
With some algebraic operations, Eq.~(\ref{eq:Hnew})  is rewritten as (see Appendix~\ref{Sec:AppH})
\begin{eqnarray}\label{eq:HY}
	\hat{\mathcal{H}}_\ell=|\mathbf{Y}_\ell |\; \vec{e}_\ell \cdot \vec{J}\;,
\end{eqnarray}
with a unit vector
\begin{widetext}
\begin{eqnarray}
	\vec{e}_\ell=\frac{1}{|\mathbf{Y}_\ell|} \left\lbrace 
	-T ( \partial_\ell \mathbf{X})+\frac{|\partial_\ell \mathbf{X}||\sin \alpha_\ell|}{|\mathbf{X}|}
	\Big\{[ \sin(T|\mathbf{X}|)-T|\mathbf{X}|] \vec{v}_{\ell,2} +\left[1-\cos(T|\mathbf{X}|) \right]  \vec{v}_{\ell,1}
	\Big\} \right\rbrace  \;,
\end{eqnarray}
where 
$\alpha_\ell:=\left<\mathbf{X},\partial_\ell \mathbf{X}\right>$ represents the angle between vectors $\mathbf{X}$ and $\partial_\ell \mathbf{X}$,  $\vec{v}_{\ell,1}=\frac{\mathbf{X}_\times \partial_\ell \mathbf{X}}{\left| \mathbf{X}\right| \left| \partial_\ell \mathbf{X}\right| \sin \alpha_\ell }$, $  \vec{v}_{\ell,2}=\frac{\mathbf{X}_{ \times 2}
	\left(\partial_\ell \mathbf{X}\right)}{\left| \mathbf{X}\right|^2 \left| \partial_\ell \mathbf{X}\right| | \sin \alpha_\ell|}$ and
\begin{eqnarray}\label{eq:Ymodul}
	|\mathbf{Y}_\ell|=\sqrt{T^2 |\partial_\ell \mathbf{X} |^2 \cos^2\alpha_\ell+\frac{4|\partial_\ell \mathbf{X}|^2 \sin^2 \alpha_\ell}{|\mathbf{X}|^2} \sin^2 \left(\frac{T|\mathbf{X}|}{2} \right)}\;.
\end{eqnarray}
\end{widetext}

For a two-dimensional pure state $\hat{\rho}_\text{in}$, its Bloch representation can be expressed by $\hat{\rho}_\text{in}=(\hat{I}+\vec{r}_\text{in} \cdot \vec{\sigma})/2=\hat{I}/2+ \vec{r}_\text{in} \cdot \vec{J}$ 
with the Bloch vector  $\vec{r}_\text{in}$ and the Pauli vector $\vec{\sigma}$. 
Moreover, we have $\text{Tr}\left[\hat{\rho}_\text{in}^2\right]=(1+|\vec{r}_\text{in}|^2)/2$ (this quantity is also referred to as  ``purity"~\cite{Purity,Paris2021properties}).
Inserting these properties and Eq.~(\ref{eq:HY})  into Eq.~(\ref{eq:QFIM_pure}), the QFI of $x_\ell$ reads
\begin{eqnarray}\label{eq:QFI_pure}
	\mathcal{F}_{\ell\ell}=|\mathbf{Y}_\ell|^2 \left[1-(\vec{e}_\ell \cdot \vec{r}_\text{in})^2\right]\;.
\end{eqnarray}
Due to $(\vec{e}_\ell \cdot \vec{r}_\text{in})^2 \ge 0$, according to Eq.~(\ref{eq:Ymodul}), the maximal QFI is deduced as
\begin{eqnarray}\label{eq:QFI_max}
\mathcal{F}^{\text {max}}_{\ell\ell}&=&|\mathbf{Y}_\ell|^2\nonumber\\
&=&T^2 |\partial_\ell \mathbf{X} |^2 \cos^2 \alpha_\ell\nonumber\\
&+&\frac{4|\partial_\ell \mathbf{X}|^2 \sin^2 \alpha_\ell}{|\mathbf{X}|^2} \sin^2 \left(\frac{T|\mathbf{X}|}{2} \right).
\end{eqnarray}
It is shown that  the first part of  $\mathcal{F}^{\text {max}}_{\ell \ell}$ is quadratic in $T$, while the second one oscillates with $T$. And the quadratic term dominantly  contributes the result of Eq.~(\ref{eq:QFI_max}) for $\alpha_\ell \neq \pi/2$. 
Eq.~(\ref{eq:QFI_max}) can  be rewritten as
\begin{eqnarray}
\mathcal{F}^{\text {max}}_{\ell\ell}\!=\! T^2 |\partial_\ell \mathbf{X} |^2 \left[\cos^2 \alpha_\ell \!+\! \sin^2 \alpha_\ell \left(\frac{\sin\left(\frac{T|\mathbf{X}|}{2}\right)}{\frac{T|\mathbf{X}|}{2}}\right)^{2} \right],\nonumber\\
\end{eqnarray}
which is bounded by a range
\begin{eqnarray}\label{eq:range}
0 \le \mathcal{F}_{\ell\ell}^\text{max} \le T^2 |\partial_\ell \mathbf{X}|^2\;.
\end{eqnarray}
In addition, the simultaneous optimal estimation of multiple parameters not only requires to maximize the QFI of each parameter, but also to ensure that all the QFI maxima are simultaneously reached.
The weak commutation condition (Eq.~(\ref{eq:wcc}))  is usually used to evaluate the second equation, which is rewritten as
\begin{eqnarray}\label{eq:wc_nc}
	\text{Tr}\left[\left[\hat{\mathcal{H}}_\ell, \hat{\mathcal{H}}_{\ell'}\right] \hat{\rho}_\text{in}\right]\!=\!\left(\frac{i}{2}\right) |\mathbf{Y}_\ell| |\mathbf{Y}_{\ell'}| (\vec{e}_\ell \times \vec{e}_{\ell'}) \cdot \vec{r}_\text{in}, 
\end{eqnarray}
for $\ell, {\ell'}={1,2,3}$ and $\ell \neq \ell'$. 
Eq.~(\ref{eq:wc_nc}) indicates that its zero value is always enslaved to the condition of $(\vec{e}_\ell \times \vec{e}_{\ell'}) \cdot \vec{r}_\text{in}=0$.

Accordingly, to achieve the simultaneous optimal estimation of $\mathbf{x}$ in the no-control case, the conditions
\begin{eqnarray}\label{eq:conditionsNC}
\left\lbrace \begin{matrix}
\vec{e}_\ell \cdot \vec{r}_\text{in}=0\;,\\
 (\vec{e}_\ell \times \vec{e}_{\ell'}) \cdot \vec{r}_\text{in}=0\;,
\end{matrix}\right.
\end{eqnarray}
need to be met at the same time for $\ell,\ell'=1,2,3$ and $\ell \neq \ell'$. 
The requirement of Eq.~(\ref{eq:conditionsNC}) cannot be satisfied simultaneously for all the parameters unless $\vec{r}_\text{in}=0$.
$\vec{r}_\text{in}=0$ means that the probe state is a maximally mixed state, i.e., $\hat{\rho}_\text{in}=\hat{I}/2$.
But the convexity and the monotonicity of the QFI manifest that the maximal QFI is attained by a pure state rather than a mixed state~\cite{PhysRevA.63.042304}. 
In the face of this contradiction, the universal solution is extending the original quantum channel to the composite quantum system~\cite{PhysRevA.63.042304}, in which a pure state can be taken as our new probe state.
Actually, this solution includes two questions: (i) how to construct this composite quantum system; (ii) how to design this pure state. 
For (i), we can introduce an ancillary channel to extend the original quantum channel. And for (ii),  we know that the reduced state of a maximally entangled two-qubit state is exactly $\hat{I}/2$.
After the rigorous calculations, we can prove that using a maximally entangled two-qubit state as the probe state and introducing an ancillary channel into the dynamics can achieve the expected simultaneous optimal estimation and remove the restriction described by Eq.~(\ref{eq:conditionsNC}). The relevant discussions are presented in Appendix~\ref{Sec:Appwc}.
\subsection{Role of quantum control}\label{Sec:withC}
As discussed in the previous works~\cite{Se, Single_Sequential_Hou, Multi_Sequential}, 
employing quantum control 
can achieve the simultaneous optimal estimation of multiple parameters in some specific systems.
The utilized sequential scheme even can provide a better estimation performance than the parallel scheme or the scheme of independently estimating each parameter~\cite{Se}.
The various differences among these three scenarios are also discussed in the configuration of quantum circuits~\cite{quintino2021deterministic} and quantum channel discrimination~\cite{PhysRevLett.127.200504}.
In this paper, we offer a different perspective for understanding the role of quantum control from the obtainable QFI result.
The QFI maximum (QFIm) is identified as a figure of merit in the following discussions.
More importantly, for the parameter $x_\ell$, we find that the promotion of the estimation precision contributed to by quantum control is not always effective. 
This dominantly depends on the characterization of an $SU(2)$ dynamics with respect to $x_\ell$, which is quantified by the angle $\alpha_\ell$.

The task on our hand now is to investigate how to simultaneously reach the upper bound of QFIm of each parameter by introducing the appropriate quantum control.
Given the form of quantum control (Eq.~(\ref{eq:Hc})), the counterpart of $\hat{\mathcal{H}}_\ell$ (Eq.~(\ref{eq:HY})) in this case can be derived by seeding Eq.~(\ref{eq:Hk}) into the computation procedures of Eqs.~(\ref{eq:partH1})-(\ref{eq:HY}). 
Accordingly, for the parameter $x_\ell$,  the resulting QFIm has no difference from Eq.~(\ref{eq:QFI_max}).
But now $\mathbf{X}$ is replaced with $\mathbf{S}$, and $\alpha_\ell$ is replaced with $\beta_\ell:=\left<\mathbf{S},\partial_\ell \mathbf{X}\right>$ that represents the angle between vectors $\mathbf{S}$ and $\partial_\ell \mathbf{X}$. 
The QFIm thus reads
\begin{eqnarray}
{\mathcal{G}}^{\text {max}}_{\ell\ell}&=&T^2 |\partial_\ell \mathbf{X} |^2 \cos^2 \beta_\ell\nonumber\\
&+&\frac{4|\partial_\ell \mathbf{X}|^2 \sin^2 \beta_\ell}{|\mathbf{S}|^2} \sin^2 \left(\frac{T|\mathbf{S}|}{2} \right)\;,
\end{eqnarray}
with $\partial_\ell \mathbf{S}=\partial_\ell \mathbf{X}$.
Particularly, with the limit of $|\mathbf{S}|\to 0$ we observe
\begin{eqnarray}\label{eq:limits}
&&\lim\limits_{|\mathbf{S}|\to 0}{\mathcal{G}}^{\text {max}}_{\ell\ell}\nonumber\\
&=&\lim\limits_{|\mathbf{S}|\to 0} T^2 |\partial_\ell \mathbf{X} |^2\left[  \cos^2 \beta_\ell
+ \sin^2 \beta_\ell\left( \frac{\sin\left(\frac{T|\mathbf{S}|}{2}\right)}{\frac{T|\mathbf{S}|}{2}}\right)^2\right] \nonumber\\
&=&T^2|\partial_\ell \mathbf{X} |^2\;,
\end{eqnarray}
which indeed arrives the upper bound of Eq.~(\ref{eq:range}).

\begin{figure}[!h]
	\centering
	\includegraphics[width=0.45\textwidth]{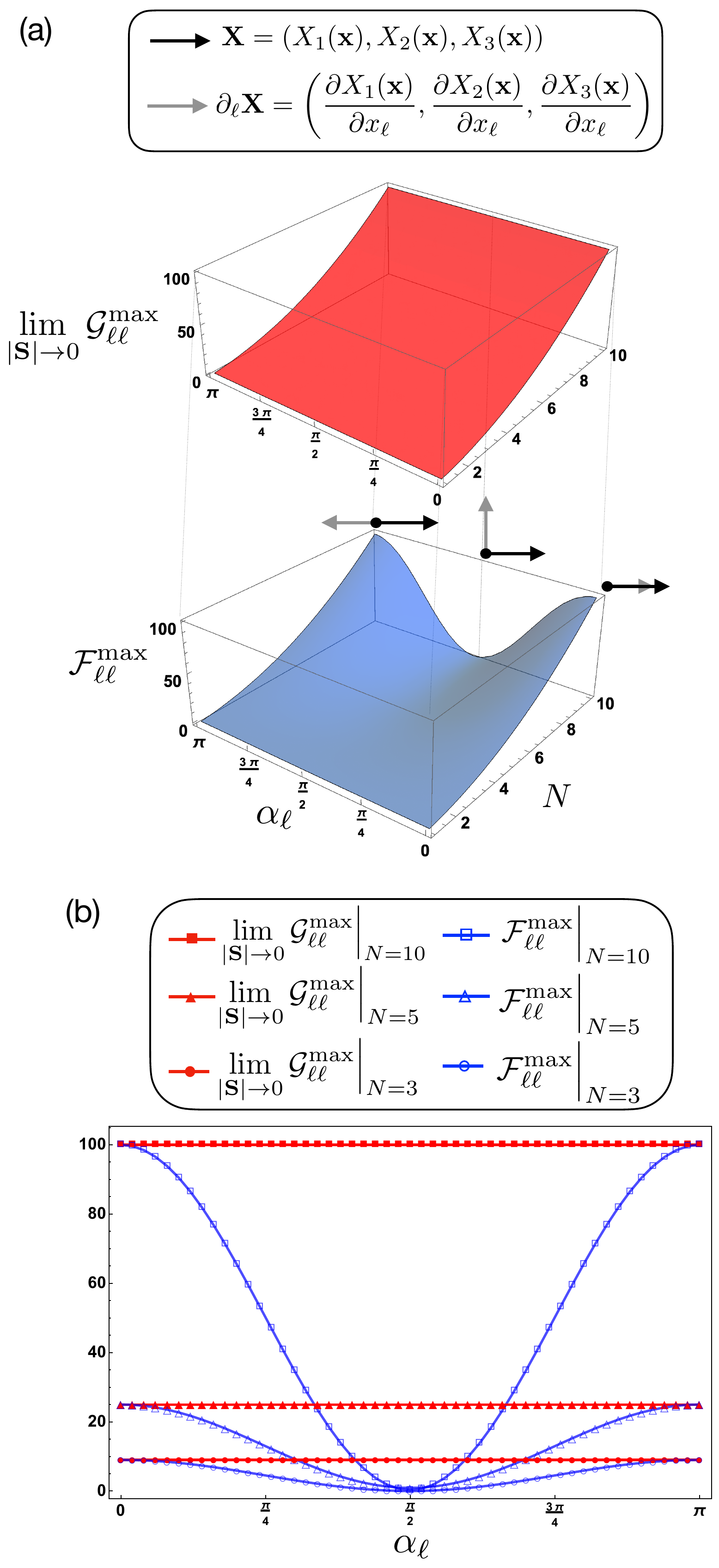}
	\caption
	{(a) The relation among $\lim_{|\mathbf{S}|\to 0}\mathcal{G}_{\ell \ell}^\text{max}$ (i.e., Eq.~(\ref{eq:limits})), $N$ and $\alpha_\ell$ is displayed by a red (upper) surface. 
		For a comparison, the relation among ${\mathcal{F}}_{\ell \ell}^\text{max}$  (i.e., Eq.~(\ref{eq:QFI_max})), $N$ and $\alpha_\ell$ is displayed by a blue (lower) surface. 
		Three typical geometrical relations between the vectors $\mathbf{X}$ and $\partial_\ell \mathbf{X}$ are also separately plotted with the given $\alpha_\ell=0,\pi/2,\pi$.
		(b) The functional dependence of Eqs.~(\ref{eq:limits}) and (\ref{eq:QFI_max}) upon $\alpha_\ell$ are respectively plotted with the given $N=3,5,10$.
		$t=1$, $|\mathbf{X}|=2$ and $|\partial_\ell \mathbf{X}|=1$ are set for the simulation.}
	\label{Fig_control}
\end{figure}

Comparing  Eq.~(\ref{eq:QFI_max}) and Eq.~(\ref{eq:limits}) we recognized that, introducing quantum control that makes $|\mathbf{S}|$ close to $0$ as far as possible corresponds to extract the quadratic term of the non-controlled QFIm and get rid of its oscialltion term.
If using $\alpha_\ell$ to quantify the characterization of an $SU(2)$ dynamics with respect to $x_\ell$, the result of Eq.~(\ref{eq:limits}) also elucidates that the quantum control makes the dependence of the estimation precision on this characterization disappear. 
The results of Eqs.~(\ref{eq:QFI_max}) and (\ref{eq:limits}) are plotted in Fig.~\ref{Fig_control} (a).
The relation among $\lim_{|\mathbf{S}|\to 0}\mathcal{G}^\text{max}_{\ell \ell}$, $N$ and $\alpha_\ell$ is displayed by a  red (upper) surface.
The relation among $\mathcal{F}^\text{max}_{\ell \ell}$, $N$ and $\alpha_\ell$ is displayed by a blue (lower) surface.
More importantly, Fig.~\ref{Fig_control} (a) exhibits that, with a given $N$, the gap between $\mathcal{F}^\text{max}_{\ell \ell}$ and $\lim_{|\mathbf{S}|\to 0}\mathcal{G}^\text{max}_{\ell \ell}$ is the largest when $\alpha_\ell=\pi/2$. 
Then we use $N=3,5,10$ as the examples to show this feature, as displayed in Fig.~\ref{Fig_control} (b).
Taking $\alpha_\ell=\pi/2$ as the axis of symmetry, the gap between $\mathcal{F}^\text{max}_{\ell \ell}$ and $\lim_{|\mathbf{S}|\to 0}\mathcal{G}^\text{max}_{\ell \ell}$ gradually decreases when $\alpha_\ell$ approaches to $0$ or $\pi$.
Up to the case where $\alpha_\ell=0$ or $\pi$, there is no gap between them.
Moreover, with a given $\alpha_\ell \in (0,\pi)$, the greater $N$ is, the more obvious their gap is. 
The physical meaning behind this phenomenon is that, for a specific type of $SU(2)$ dynamics in which $\mathbf{X}$ is perpendicular to $\partial_\ell \mathbf{X}$, the promotion of $\mathcal{F}^\text{max}_{\ell \ell}$ can get the most benefits from quantum control. 
Then the contribution of quantum control on the precision improvement of $x_\ell$ gradually reduces when $\mathbf{X}$ and $\partial_\ell \mathbf{X}$ tend to be colinear. 
And for a different class of $SU(2)$ dynamics in which $\mathbf{X}$ and $\partial_\ell \mathbf{X}$ are completely colinear,  the controls do not play a role in the precision enhancement of $x_\ell$. 
It follows that the promotion of the estimation precision contributed by quantum control depends on the characterization of an $SU(2)$ dynamics with respect to the objective parameter.

Furthermore, limiting $|\mathbf{S}| \to 0$ corresponds to using a quantum control designed as
\begin{eqnarray}\label{eq:XcM}
	\mathbf{X}_c=-\mathbf{X}\;,
\end{eqnarray}
in our research framework.
Repeating the computation procedures of Eqs.~(\ref{eq:partH1})-(\ref{eq:HY}) with Eq.~(\ref{eq:XcM}), 
the counterpart of Eq.~(\ref{eq:HY}) is
\begin{eqnarray}\label{eq:H_with}
\widetilde{\mathcal{H}}_\ell=-T\left( \partial_\ell \mathbf{X} \cdot \vec{J}\right)\;.
\end{eqnarray}
Inserting Eq.~(\ref{eq:H_with}) into Eq.~(\ref{eq:QFIM_pure}), the control-aided QFI  of  $x_\ell$ reads
\begin{eqnarray}
\widetilde{\mathcal{F}}_{\ell\ell}=T^2 \left[|\partial_\ell \mathbf{X}|^2-(\partial_\ell \mathbf{X} \cdot \vec{r}_\text{in})^2 \right]\;.
\end{eqnarray}
Due to $\left(\partial_\ell \mathbf{X} \cdot \vec{r}_\text{in}\right)^2 \ge 0$, the QFIm is
\begin{eqnarray}\label{eq:QFImWC}
\widetilde{\mathcal{F}}^{\text {max}}_{\ell\ell}=T^2 |\partial_\ell \mathbf{X}|^2\;.
\end{eqnarray}
Obviously, $\widetilde{\mathcal{F}}^{\text {max}}_{\ell\ell}=\lim_{|\mathbf{S}|\to 0}{\mathcal{G}}^{\text {max}}_{\ell\ell}$.
Our design for quantum control (Eq.~(\ref{eq:XcM})) also coincides with the previous work~\cite{Se,Multi_Sequential}.
Next we consider whether all $\widetilde{\mathcal{F}}^{\text {max}}_{\ell\ell}$ for $\ell=1,2,3$ can be simultaneously attained. 
The weak commutation condition (Eq.~(\ref{eq:wcc})) is renewed as
\begin{eqnarray}\label{eq:wcrnew}
	\text{Tr}\left[\left[\widetilde{\mathcal{H}}_\ell, \widetilde{\mathcal{H}}_{\ell'}\right] \hat{\rho}_\text{in}\right] \!=\! \left(\frac{i}{2} \right) T^2 (\partial_\ell \mathbf{X} \times \partial_{\ell'} \mathbf{X})\cdot \vec{r}_\text{in}\;,
\end{eqnarray}
for  $\ell, \ell'={1,2,3}$ and $\ell\neq \ell'$. 
We observe that the result of Eq.~(\ref{eq:wcrnew}) equals to zero if and only if $(\partial_\ell \mathbf{X} \times \partial_{\ell'} \mathbf{X})\cdot \vec{r}_\text{in}=0$.

Therefore, for achieving the simultaneous optimal estimation of $\mathbf{x}$ in the control-aided case, the conditions 
\begin{eqnarray}\label{eq:twoconditions}
	\left\lbrace \begin{matrix}
		\partial_\ell \mathbf{X} \cdot \vec{r}_{\text{in}}=0\;,\\
		(\partial_\ell \mathbf{X} \times \partial_{\ell'} \mathbf{X})\cdot \vec{r}_\text{in}=0\;,
	\end{matrix}\right.
\end{eqnarray}
need to be met at the same time for $\ell,\ell'=1,2,3$ and $\ell \neq \ell'$. 
Similar to Eq.~(\ref{eq:conditionsNC}), Eq.~(\ref{eq:twoconditions}) encounters a similar dilemma that the required conditions cannot be simultaneously satisfied unless $\vec{r}_\text{in}=0$. So the previous analyses for Eq.~(\ref{eq:conditionsNC}) also apply to the current case.
We can use a maximally entangled two-qubit state as the probe state and introduce an ancillary channel into the dynamics to jointly achieve the desired simultaneous optimal estimation. 
The pertinent details are presented in Sec.~\ref{Sec:Ancillary}.
Up to now we have given a set of QFI results and the corresponding weak commutation conditions, and clarified the effectiveness of quantum control in promoting multiparameter estimation precision.
Sequentially, we place an emphasis on a pivotal problem, i.e., the attainability of the control-enhanced simultaneous optimal estimation.

\section{Attainability of the control-enhanced simultaneous optimal estimation}\label{Sec:Ancillary}
To achieve the optimal quantum estimation with respect to all the parameters, there are two steps that should be followed:
(i) maximizing the QFI of each parameter;
(ii) simultaneously achieving all the QFI maxima.
In Sec.~\ref{Sec:withC} where a two-dimensional pure state is used to be the probe state, we find that these two steps bring a set of restriction conditions, as expressed by Eq.~(\ref{eq:twoconditions}).
Moreover, these conditions generally cannot be satisfied unless designing the probe state as $\hat{\rho}_\text{in}=\hat{I}/2$ (i.e., the Bloch vector $\vec{r}_\text{in}$ is a zero vector). This exactly corresponds to the reduced state of a maximally entangled two-qubit state. 
Inspired from this point, we can prove that the restriction induced by Eq.~(\ref{eq:twoconditions}) can be removed by virtue of a maximally entangled  state and an ancillary channel.

As shown in Fig.~\ref{Fig_scheme} (b), an extra ancillary channel is configured in the scenario but it does not interact with the dynamics.
For the moment, we consider the simplest case that only includes one ancillary channel.
The total unitary operator is updated as
\begin{eqnarray}\label{eq:new_totalU}
	\hat{U}_{tot,A}&=&\left((e^{-it\hat{H}_c} \otimes \hat{I}_A )(e^{-it\hat{H}(\mathbf{x})} \otimes \hat{I}_A)\right)^N\nonumber\\
	&=&\left(e^{-it(\hat{H}(\mathbf{x})+\hat{H}_c)}\right)^N \otimes \hat{I}_A\;,
\end{eqnarray}
where $\hat{I}_A$ is the identity operator.
Combining  Eq.~(\ref{eq:new_totalU}) with Eq.~(\ref{eq:partH1}) and repeating the calculation procedures presented in Sec.~\ref{Sec:withC}, Eq.~(\ref{eq:H_with}) is rewritten as
\begin{eqnarray}\label{eq:H1}
	\widetilde{\mathcal{H}}_{\ell,A}=-T \left(\partial_\ell \mathbf{X} \cdot \vec{J}\right)\otimes \hat{I}_A\;,
\end{eqnarray}
which indicates
\begin{eqnarray}
	[\widetilde{\mathcal{H}}_{\ell,A},\widetilde{\mathcal{H}}_{\ell',A}]=iT^2 (\partial_\ell \mathbf{X} \times \partial_{\ell'} \mathbf{X}) \cdot \vec{J} \otimes \hat{I}_A\;.
\end{eqnarray}
Then taking a pure state $|\psi_{SA}\rangle$ to seed Eq.~(\ref{eq:wcc}), we have
\begin{eqnarray}\label{eq:Arelation}
	&&\text{Tr}\left[\hat{\rho}_\text{in} [\widetilde{\mathcal{H}}_{\ell,A},\widetilde{\mathcal{H}}_{{\ell'},A}]\right]\nonumber\\
	&=&\text{Tr}\left[ \hat{\rho}_\text{in}
	\left(iT^2 (\partial_\ell \mathbf{X} \times \partial_{\ell'} \mathbf{X}) \cdot \vec{J} \otimes \hat{I}_A\right)\right]\nonumber\\
	&=&\text{Tr}\left[\hat{\rho}_S \left(iT^2 (\partial_\ell \mathbf{X} \times \partial_{\ell'} \mathbf{X}) \cdot \vec{J}\right)\right]\;,
\end{eqnarray}
where $\hat{\rho}_\text{in}=|\psi_{SA}\rangle \langle \psi_{SA}|$ and $\hat{\rho}_S$ is the reduced density operator of $\hat{\rho}_\text{in}$ after tracing out the ancillary part, i.e., $\hat{\rho}_S=\text{Tr}_A[\hat{\rho}_\text{in}]$.
Sequentially, we diagonalize  the matrix  $iT^2 (\partial_\ell \mathbf{X} \times \partial_{\ell'} \mathbf{X}) \cdot \vec{J}$ as
\begin{eqnarray}\label{eq:newdia}
&&iT^2 (\partial_\ell \mathbf{X} \times \partial_{\ell'} \mathbf{X}) \cdot \vec{J}\nonumber\\
&=&\hat{W}\!\!
	\left( \begin{matrix}
		+c T^2 |\partial_\ell \mathbf{X} \times \partial_{\ell'} \mathbf{X}|&0\\\\
		0&-c T^2 |\partial_\ell \mathbf{X} \times \partial_{\ell'} \mathbf{X}|
	\end{matrix}\right)\!\!
	\hat{W}^\dagger\;, \nonumber\\
\end{eqnarray}
where $\hat{W}$ is a unitary matrix. $+c$ and $-c$  separately represent the maximal and  minimal eigenvalues of $\hat{j}_m$ (the generator of $SU(2)$ algebra) for $m =1,2,3$.
The new quantum state $\hat{\rho}^\prime$ is defined as $\hat{\rho}^\prime=\hat{W}^\dagger \hat{\rho}_S \hat{W}$ with the diagonal elements ${\rho_{11}^\prime}$, ${\rho_{22}^\prime}$.
Inserting Eq.~(\ref{eq:newdia}) and $\hat{\rho}{'}$ into Eq.~(\ref{eq:Arelation}), we have
\begin{eqnarray}\label{eq:Awakcrelation}
	&&\text{Tr}\left[\hat{\rho}_\text{in} [\tilde{\mathcal{H}}_{\ell,A},\tilde{\mathcal{H}}_{{\ell'},A}]  \right]\nonumber\\
	&=&\text{Tr}\left[\hat{\rho}^\prime \left( \begin{matrix}
		+c T^2 |\partial_\ell \mathbf{X} \times \partial_{\ell'} \mathbf{X}|&0\\\\
		0&-c T^2 |\partial_\ell \mathbf{X} \times \partial_{\ell'} \mathbf{X}|
	\end{matrix}\right) \right]\nonumber\\
	&=&({\rho}^\prime_{11}-{\rho}^\prime_{22}) c T^2 |\partial_\ell \mathbf{X} \times \partial_{\ell'} \mathbf{X}|\;,
\end{eqnarray}
for $\ell \neq \ell'$. If $\hat{\rho}^\prime=\hat{I}/2$ the result of Eq.~(\ref{eq:Awakcrelation}) equals to zero, which also
means $\hat{\rho}_S=\hat{I}/2$ so that $\hat{\rho}_\text{in}$ is a maximally entangled two-qubit state.
Accordingly, we can see that as long as using a maximally entangled two-qubit state and an ancillary channel, the weak commutation condition can be unconditionally met. 
This consequence amounts to eliminating the restriction caused by Eq.~(\ref{eq:twoconditions}).
We also get a similar conclusion for the no-control case in Appendix~\ref{Sec:Appwc}.
In addition, the employed maximally entangled two-qubit state can unconditionally achieve the same QFIm with the two-dimensional pure state satisfying $\partial_\ell \mathbf{X} \cdot \vec{r}_\text{in}=0$, for which Appendix~\ref{Sec:Appsame} gives the demonstration.

We worked out the QCRBs for the parameters and shown that these parameters can simultaneously get their QCRBs when the probe state is the maximally entangled state. However, to saturate the QCRBs, one needs to choose the optimal measurement scheme ~\cite{Se,Multi_Sequential}.

\section{Example: estimating an unknown magnetic field in a spin-$1/2$ system}\label{Sec:Example}
Now we consider a nontrivial physical scene where a spin-$1/2$ particle (such as an electron) is placed in a magnetic field.
Three to-be-estimated parameters are $\mathbf{x}:=\{B,\theta,\phi\}$ denoting the magnitude and the direction of the magnetic field.
This canonical system has also been investigated in previous works with the different analytical methods~\cite{Se,Multi_Sequential}.
The Hamiltonian is $\hat{H}(\mathbf{x})=B \vec{n}_0 \cdot \vec{\sigma}=2B \vec{n}_0 \cdot \vec{J}$, in which $\vec{n}_0=(\sin \theta \cos \phi, \sin \theta \sin \phi, \cos \theta)$ and  $\vec{\sigma}=(\hat{\sigma}_x,\hat{\sigma}_y,\hat{\sigma}_z)$ represents the three-component vector of Pauli matrices.
According to Eq.~(\ref{eq:HSU(2)}), we have
\begin{eqnarray}
\mathbf{X}=2B\vec{n}_0,\;
\partial_B \mathbf{X}=2\vec{n}_0, \;
\partial_\theta \mathbf{X}=2B\vec{n}^{'}_0,\;
\partial_\phi \mathbf{X}=2B\vec{n}^{''}_0, \nonumber\\
\end{eqnarray}
with $\vec{n}^{'}_0=(\cos\theta \cos\phi, \cos\theta \sin\phi,-\sin\theta)$ and $\vec{n}^{''}_0=(-\sin\theta \sin\phi, \sin\theta \cos\phi, 0)$.
The $SU(2)$ dynamics with respect to three parameters can be characterized by three angles
$\alpha_B=\left<\mathbf{X}, \partial_B \mathbf{X}\right>=0$,
$\alpha_\theta=\left<\mathbf{X}, \partial_\theta \mathbf{X}\right>={\pi}/{2}$ and $\alpha_\phi=\left<\mathbf{X}, \partial_\phi \mathbf{X}\right>={\pi}/{2}$.	
These three angles, according to the discussions in Sec.~\ref{Sec:withC}, indicate that the promotion of the estimation precision with respect to $\{\theta,\phi\}$ can get the most benefits from the controls.
On the contrary, the controls will lose their advantages when $B$ is estimated. 
The following calculation results can prove these verdicts.
\subsubsection{Without controls}\label{Subsec:without}
In the absence of quantum control (i.e., $\mathbf{X}_c$ is a zero vector), according to Eq.~(\ref{eq:HY}) we have
\begin{eqnarray}\label{eq:HB}
\hat{\mathcal{H}}_B
=|\mathbf{Y}_B|\; \vec{e}_B \cdot \vec{J}\;,
\end{eqnarray}
with
\begin{eqnarray}
	|\mathbf{Y}_B|&=&2T\;,\\
	\vec{e}_B&=&\left(-\sin\theta\cos\phi,-\sin\theta \sin\phi,-\cos\theta\right) \;,
\end{eqnarray}
where $\vec{e}_B=-\vec{n}_0$.
Inserting Eq.~(\ref{eq:HB}) into Eq.~(\ref{eq:QFIM_pure}), the QFI of $B$ is 
\begin{eqnarray}\label{B}
\mathcal{F}_B=4T^2\left(1-(\vec{e}_B \cdot \vec{r}_{\text{in}})^2\right)\;.
\end{eqnarray}
Since $(\vec{e}_B \cdot \vec{r}_{\text{in}})^2 \ge 0$, the maximum of Eq.~(\ref{B}) is
\begin{eqnarray}\label{eq:QFIBWC}
\mathcal{F}_{B}^{\text{max}}=4T^2\;.
\end{eqnarray} 
Similarly, for the parameter $\theta$ we obtain
\begin{eqnarray}\label{eq:Htheta}
\hat{\mathcal{H}}_\theta
=|\mathbf{Y}_\theta|\; \vec{e}_\theta \cdot \vec{J}\;,
\end{eqnarray}
with
\begin{eqnarray}
|\mathbf{Y}_\theta|&=&2\sin(BT)\;,\\
\vec{e}_\theta&=&(-\cos(BT)\cos\theta\cos\phi-\sin(BT)\sin\phi,\nonumber\\
&&\cos\phi \!-\! \cos(BT)\cos\theta\sin\phi, \cos(BT)\sin\theta).
\end{eqnarray}
Substituting Eq.~(\ref{eq:Htheta}) into Eq.~(\ref{eq:QFIM_pure}), the QFI of $\theta$ is
\begin{eqnarray}
	\mathcal{F}_\theta=4\sin^2(BT)\left(1-(\vec{e}_\theta \cdot \vec{r}_{\text{in}})^2\right)\;.
\end{eqnarray}
According to $(\vec{e}_\theta \cdot \vec{r}_{\text{in}})^2 \ge 0$,  the maximal $\mathcal{F}_\theta$ is 
\begin{eqnarray}\label{eq:maxtheta}
	\mathcal{F}^{\text{max}}_\theta=4\sin^2 (BT)\;.
\end{eqnarray}
For the parameter $\phi$, we likewise get
\begin{eqnarray}\label{eq:H_phi}
\hat{\mathcal{H}}_\phi
=|\mathbf{Y}_\phi|\; \vec{e}_\phi \cdot \vec{J}\;,
\end{eqnarray}
with 
\begin{eqnarray}
|\mathbf{Y}_\phi|&\!=\!&2\sin(BT)\sin\theta\;,\\
\vec{e}_\phi&=&(-\cos\theta \cos \phi \sin(BT)+\cos(BT)\sin \phi,\nonumber\\
&&\!-\! \cos(BT)\cos\phi \!-\! \sin(BT)\cos\theta \sin \phi, \nonumber\\
&& \sin(BT)\sin\theta).
\end{eqnarray}
Combining Eq.~(\ref{eq:H_phi}) with Eq.~(\ref{eq:QFIM_pure}), the QFI of $\phi$ reads
\begin{eqnarray}
\mathcal{F}_\phi=4 \sin^2 \theta \sin^2 (BT) \left(1-(\vec{e}_\phi \cdot \vec{r}_{\text{{in}}})^2\right)\;,
\end{eqnarray}
its maximum is 
\begin{eqnarray}
	\mathcal{F}^{\text{max}}_\phi=4\sin^2 \theta\sin^2 (BT)\;,
\end{eqnarray}
as $(\vec{e}_\phi \cdot \vec{r}_{\text{{in}}})^2 \ge 0$.
Apart from the diagonal elements of the QFIM, the off-diagonal elements are also investigated.
If all the parameters reach to individual highest estimation precision, the off-diagonal entries are totally zeros (see Appendix~\ref{Sec:APPoff}).
Therefore, after executing the $SU(2)$ coding unitary evolution $N$ times without any control operation, the optimal QFIM reads
\begin{eqnarray}\label{eq:MaxQFIM}
\mathcal{F}^\text{max}=\left(  \begin{matrix}
4T^2&0&0\\
0&4\sin^2(BT)&0\\
0&0&4 \sin^2(BT)\sin^2\theta
\end{matrix}\right) \;.
\end{eqnarray}
Combining Eq.~(\ref{tmp1}) and Eq.~(\ref{eq:MaxQFIM}), we notice that 
the highest estimation precision (the minimal standard deviation) of $\theta$ and $\phi$ are always bounded with the increase of $T$. 
Their QFI results only involve the oscillation term without the quadratic term (see the analyses for Eq.~(\ref{eq:QFI_max})), the obtainable estimation precision is rather limited.
In this case, the weak commutation condition (Eq.~(\ref{eq:wcc})) is specified as
\begin{eqnarray}
	\text{Tr}\left[ \left[\hat{\mathcal{H}}_B,\hat{\mathcal{H}}_\theta \right] \hat{\rho}_\text{in}
\right]&=&-2iT\sin(BT)\vec{e}_\phi \cdot \vec{r}_\text{in}\;, \label{eq:weak1}  \\
	\text{Tr}\left[ \left[\hat{\mathcal{H}}_B,\hat{\mathcal{H}}_\phi \right] \hat{\rho}_\text{in}
\right]&=&2iT\sin\theta \sin(BT) \vec{e}_\theta \cdot \vec{r}_\text{in}\;, \label{eq:weak2}\\
	\text{Tr}\left[ \left[\hat{\mathcal{H}}_\theta,\hat{\mathcal{H}}_\phi \right] \hat{\rho}_\text{in}
\right]&=&-2i\sin\theta \sin^2 (BT)\vec{e}_B \cdot \vec{r}_\text{in}\;,	\label{eq:weak3}
\end{eqnarray}	
where $\vec{e}_B\times\vec{e}_\theta=-\vec{e}_\phi$, $\vec{e}_B \times \vec{e}_\phi=\vec{e}_\theta$ and $\vec{e}_\theta \times \vec{e}_\phi=-\vec{e}_B$ are used.

\begin{figure}[!h]
	\centering
	\includegraphics[width=0.45\textwidth]{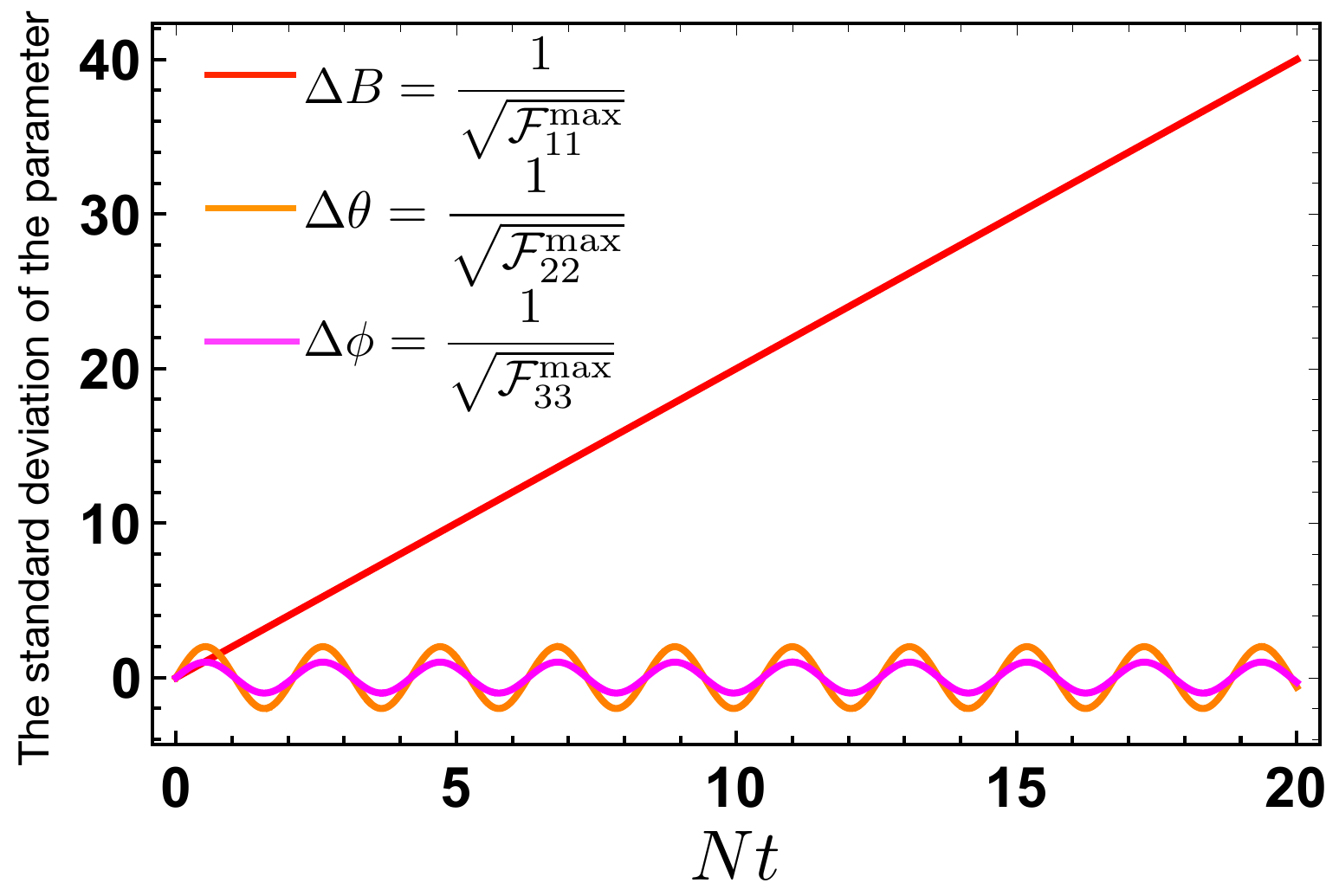}
	\caption
	{In the scheme of Fig.~\ref{Fig_scheme} (b) where quantum control is characterized by $\mathbf{X}_c=\vec{0}$ (zero vector), the highest estimation precision expressed by the standard deviation for $B,\theta$ and $\phi$  are plotted separately. 
	All the relevant parameters are rescaled by $t$, and $t=1$, $Bt=3$ and $\theta t=\pi/6$.}
	\label{Fig_simwithout}
\end{figure}

To achieve the simultaneous optimal estimation for $B,\theta$ and $\phi$, the result of Eq.~(\ref{eq:MaxQFIM}) and the zero values of Eqs.~(\ref{eq:weak1})-(\ref{eq:weak3}) need to be reached at the same time. 
It follows that the condition
\begin{eqnarray}\label{eq:example_condition}
\vec{e}_B \cdot \vec{r}_{\text{in}}=\vec{e}_\theta \cdot \vec{r}_{\text{in}}=\vec{e}_\phi \cdot \vec{r}_{\text{in}}=0\;,
\end{eqnarray}
should be met. 
Eq.~(\ref{eq:example_condition}) is a specific example of Eq.~(\ref{eq:conditionsNC}).
Due to $\vec{e}_B$, $\vec{e}_\theta$ and $\vec{e}_\phi$ are perpendicular to each other, we cannot find a nonzero $\vec{r}_\text{in}$ that is perpendicular to each of them. 
But  this difficulty can be overcome if using a maximally entangled two-qubit state as the probe state and introducing an ancillary channel into the configuration (as discussed in Appendix~\ref{Sec:Appwc}).
In such a case, the attainable highest estimation precision $\Delta B$, $\Delta \theta$ and $\Delta \phi$ can be figured out by substituting the diagonal elements of Eq.~(\ref{eq:MaxQFIM}) into Eq.~(\ref{tmp1}) where $M$ is set to be 1 for simplicity. We plot the results in Fig.~\ref{Fig_simwithout} where $\Delta \theta$ and $\Delta \phi$ are always bounded with the increase of the total evolution time $T=Nt$. 
\subsubsection{With controls}\label{Subsec:with}
In the presence of quantum control and $\mathbf{X}_c=-\mathbf{X}$, according to Eq.~(\ref{eq:H_with}) we obtain
\begin{eqnarray}\label{eq:HBnew}
	\widetilde{\mathcal{H}}_B=-T \partial_B \left(2B \vec{n}_0 \cdot \vec{J}\right)=-2T \vec{n}_0 \cdot \vec{J}\;,
\end{eqnarray}
which is consistent with Eq.~(\ref{eq:HB}).
Inserting Eq.~(\ref{eq:HBnew}) into Eq.~(\ref{eq:QFIM_pure}) we have
\begin{eqnarray}
\widetilde{\mathcal{F}}_B=4T^2\left(1-(\vec{n}_0 \cdot \vec{r}_{\text{in}})^2\right)\;.
\end{eqnarray}
The maximal $\widetilde{\mathcal{F}}_B$ is
\begin{eqnarray}\label{eq:QFIB}
\widetilde{\mathcal{F}}_B^{\text{max}}=4T^2\;.
\end{eqnarray} 
As we clarified in the beginning, this dynamics makes $\alpha_B=0$ such that the controls do not play a role in improving the estimation precision of $B$. 
The same QFI maxima (Eq.~(\ref{eq:QFIBWC}) and Eq.~(\ref{eq:QFIB})) indeed prove this point. 
Then for the parameter $\theta$ we have
\begin{eqnarray}\label{eq:Hthetanew}
	\widetilde{\mathcal{H}}_\theta=-T \partial_\theta \left(2B \vec{n}_0 \cdot \vec{J}\right)=-2TB \vec{n}^{'}_0 \cdot \vec{J}\;,
\end{eqnarray}
which is plugged into Eq.~(\ref{eq:QFIM_pure}), we finally get
\begin{eqnarray}\label{rrr}
\widetilde{\mathcal{F}}_\theta=4(BT)^2 (1-(\vec{n}^{'}_0 \cdot \vec{r}_\text{in})^2)\;.
\end{eqnarray}
Since $(\vec{n}^{'}_0 \cdot \vec{r}_\text{in})^2 \ge 0$, the maximum of Eq.~(\ref{rrr}) is
\begin{eqnarray}
\widetilde{\mathcal{F}}_\theta^\text{max}=4(BT)^2\;.
\end{eqnarray}
Similarly, for the parameter $\phi$ we acquire
\begin{eqnarray}\label{eq:Hphinew}
\widetilde{\mathcal{H}}_\phi=-T \partial_\phi \left(2B \vec{n}_0 \cdot \vec{J}\right)=-2TB \vec{n}^{''}_0 \cdot \vec{J}\;.
\end{eqnarray}
The resulting QFI  reads
\begin{eqnarray}
\widetilde{\mathcal{F}}_\phi=4(BT)^2 (\sin^2\theta-(\vec{n}^{''}_0 \cdot \vec{r}_\text{in})^2)\;,
\end{eqnarray}
the corresponding maximum is
\begin{eqnarray}
\widetilde{\mathcal{F}}^\text {max}_\phi=4(BT)^2 \sin^2 \theta\;,
\end{eqnarray}
as $(\vec{n}^{''}_0 \cdot \vec{r}_\text{in})^2 \ge 0$.
The off-diagonal entries of this QFIM are also studied, which are totally zeros when all the parameters reach to individual highest estimation precision (see Appendix~\ref{Sec:APPoff}).
Accordingly, the QFIM optimized by $N$ controls is
\begin{eqnarray}\label{eq:MaxQFIMnew}
\widetilde{\mathcal{F}}^\text{max}=\left(  \begin{matrix}
		4T^2&0&0\\
		0&4(BT)^2&0\\
		0&0&4 (BT)^2\sin^2\theta
	\end{matrix}\right) \;.
\end{eqnarray}
Compared with  Eq.~(\ref{eq:MaxQFIM}), Eq.~(\ref{eq:MaxQFIMnew}) manifests that the highest estimation precision (the minimal standard deviation) of $\theta$ and $\phi$ both are significantly promoted with the increase of $T$. 
Their QFI results only keep the quadratic terms (see the discussions for Eq.~(\ref{eq:QFI_max})), the estimation precision  thus can be notably enhanced.
One can write out the weak commutation condition (Eq.~(\ref{eq:wcc})) as
\begin{eqnarray}
	\text{Tr}\left[ \left[\tilde{\mathcal{H}}_B,\tilde{\mathcal{H}}_\theta \right] \hat{\rho}_\text{in}\right]&=&\frac{2iT^2 B}{\sin \theta} \vec{n}^{''}_0 \cdot \vec{r}_\text{in}\;,\label{eq:C1}\\
	\text{Tr}\left[ \left[\tilde{\mathcal{H}}_B,\tilde{\mathcal{H}}_\phi \right] \hat{\rho}_\text{in}\right]&=&-2iT^2 B \sin\theta \vec{n}^{'}_0 \cdot \vec{r}_\text{in}\;,\label{eq:C2}\\
	\text{Tr}\left[ \left[\tilde{\mathcal{H}}_\theta,\tilde{\mathcal{H}}_\phi \right] \hat{\rho}_\text{in}\right]&=&2iT^2 B^2 \sin \theta \vec{n}_0 \cdot \vec{r}_\text{in}\;,\label{eq:C3}
\end{eqnarray}
where $\vec{n}_0 \times \vec{n}^{'}_0=\vec{n}^{''}_0/\sin\theta$, $\vec{n}_0 \times \vec{n}^{''}_0=-\sin\theta \vec{n}^{'}_0$ and $\vec{n}^{'}_0 \times \vec{n}^{''}_0=\sin \theta \vec{n}_0$ are used.

\begin{figure}[!h]
	\centering
	\includegraphics[width=0.45\textwidth]{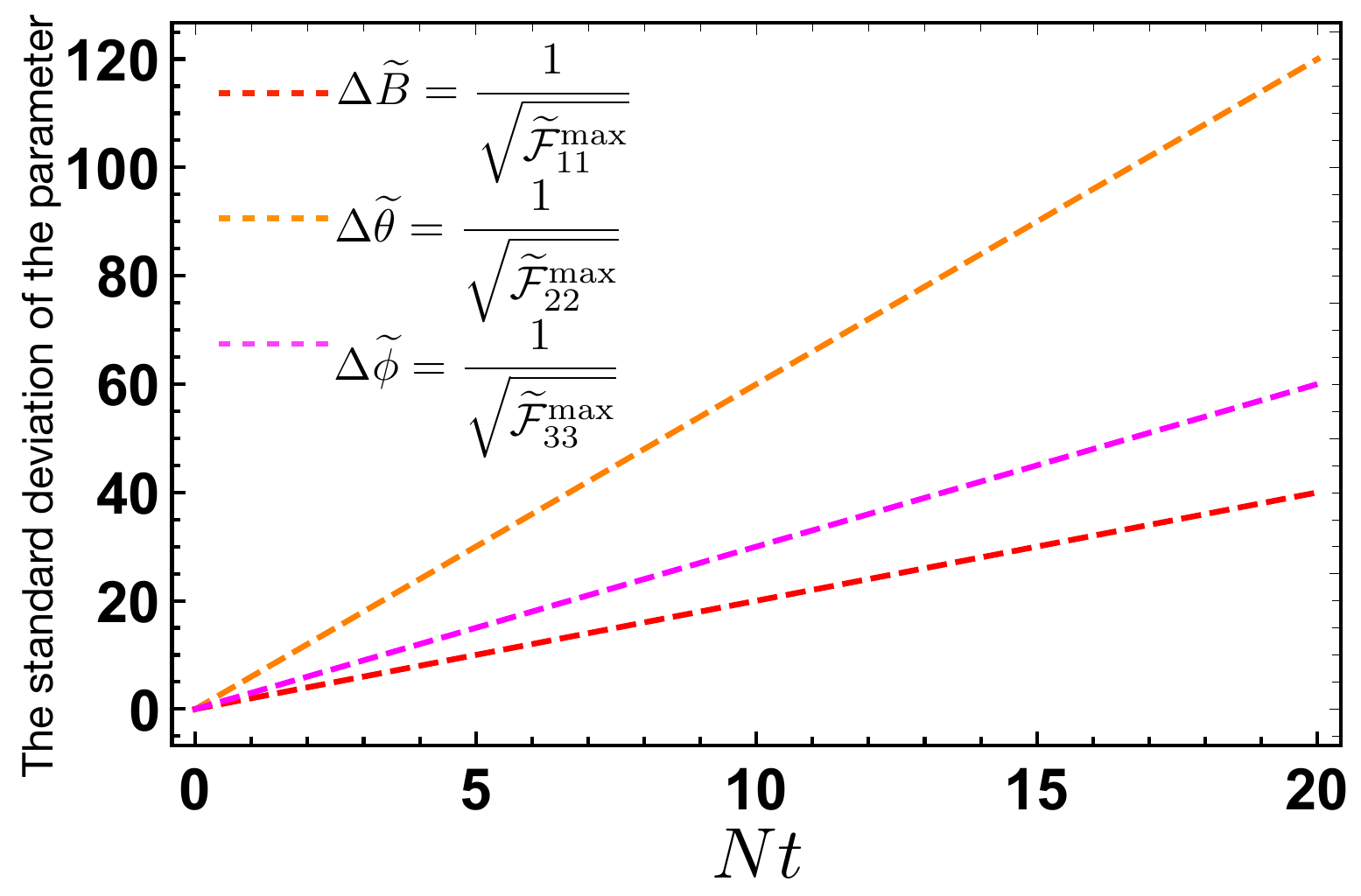}
	\caption
	{In the scheme of Fig.~\ref{Fig_scheme} (b) where quantum control is designed as $\mathbf{X}_c=-\mathbf{X}$, the highest estimation precision expressed by the standard deviation for $B,\theta$ and $\phi$ are plotted separately. 
	All the relevant parameters are rescaled by $t$, and $t=1$, $Bt=3$ and $\theta t=\pi/6$.}
	\label{Fig_simwith}
\end{figure}

With the help of quantum control, the simultaneous optimal estimation for $B,\theta$ and $\phi$ requires that the result of Eq.~(\ref{eq:MaxQFIMnew}) and the zero values of Eqs.~(\ref{eq:C1})-(\ref{eq:C3}) are reached simultaneously. 
Accordingly,  the condition
\begin{eqnarray}\label{eq:condition}
\vec{n}_0 \cdot \vec{r}_\text{in}=\vec{n}^{'}_0 \cdot \vec{r}_\text{in}=\vec{n}^{''}_0 \cdot \vec{r}_\text{in}=0\;,
\end{eqnarray}
need to be satisfied.
Eq.~(\ref{eq:condition}) is a specific example of Eq.~(\ref{eq:twoconditions}).
But now we encounter a barrier like in the no-control scheme.
Due to $\vec{n}_0$, $\vec{n}^{'}_0$ and $\vec{n}^{''}_0$ are perpendicular to each other, it is infeasible to find a nonzero $\vec{r}_\text{in}$ that is perpendicular to each of them.
Fortunately, using a maximally entangled two-qubit state as the probe state and adding an ancillary channel in the configuration, this barrier can be removed (as discussed in Sec.~\ref{Sec:Ancillary}).
By virtue of these solutions, the reachable highest estimation precision $\Delta \widetilde{B}$, $\Delta \widetilde{\theta}$ and $\Delta \widetilde{\phi}$ can be determined by substituting the diagonal elements of Eq.~(\ref{eq:MaxQFIMnew}) into Eq.~(\ref{tmp1}) where $M$ is set to be 1 for simplicity.
Their results are plotted in Fig.~\ref{Fig_simwith}, in which $\Delta \widetilde{B}$, $\Delta \widetilde{\theta}$ and $\Delta \widetilde{\phi}$ can be linearly promoted with the increase of the total evolution time $T=Nt$. 
The Heisenberg scalings $1/T$ for $B,\theta$ and $\phi$ are thus fulfilled simultaneously. 

\section{Conclusion}\label{Sec:Conclusion}
We developed a method associated with $\mathbf{X}$  and its partial derivative $\partial_\ell \mathbf{X}$,  which is employed to investigate the control-enhanced quantum multiparameter estimation with an $SU(2)$ coding unitary evolution. 
Based on working out the QFI maximum of each parameter, we  revealed that quantum control is not always effective in improving the estimation precision of multiple parameters, which largely depends on the characterization of an $SU(2)$ dynamics with respect to the objective parameter (quantified by the angle $\alpha_\ell$ between the vectors  $\mathbf{X}$ and  $\partial_\ell \mathbf{X}$).
For an $SU(2)$ dynamics with $\alpha_\ell=\pi/2$, the ability of quantum control to enhance the estimation precision can play the greatest advantage. In contrast, for an $SU(2)$  dynamics characterized by $\alpha_\ell=0$ or $\pi$, quantum control loses its power.
In addition, a set of conditions restricting the simultaneous optimal estimation of all the parameters were presented. Then we  proved that using the maximally entangled two-qubit state and an ancillary channel can remove these restrictions.
A spin-$1/2$ system used to estimate the attributes of an unknown magnetic field was investigated at the last as an example, the relevant results were also summarized in two tables of the Appendix~\ref{App:tables}.

The Hamiltonian discussed in this paper is time independent, if it is time dependent, for instance estimating the magnitude and the frequency of a rotating magnetic field given by $\hat{H}(\mathbf{x})=-2B \vec{n}_\xi \cdot \vec{J}$ with $\vec{n}_\xi=(\cos(\omega t),0,\sin(\omega t))$, the form of quantum control needs to be correspondingly altered~\cite{pang2017optimal,Hourotate}. 
In addition, in the experiment, the employed $\mathbf{X}_c$ (Eq.~(\ref{eq:XcM})) relates to the ture value of $\mathbf{x}$ that is however unknown in advance. Hence, the form of the control has to be adaptively updated as $-\mathbf{X}(\tilde{\mathbf{x}})$ where $\tilde{\mathbf{x}}$ represents a set of estimated values of $\mathbf{x}$.
When $\tilde{\mathbf{x}}=\mathbf{x}$, the employed $\mathbf{X}_c$ is the expected optimal quantum control.
It follows that quantum control used in our scheme is essentially a kind of  ``adaptively feedback control"~\cite{Fujiwara_2006}.
Then how to make the temporary control converges to the optimal form as fast as possible, is the subsequent problem. 
Some algorithm such as the gradient ascent pulse engineering (GRAPE)~\cite{PhysRevA.96.042114}, the Krotov algorithm~\cite{reich2012monotonically,PhysRevRes}, the chopped random-basis (CRAB) method~\cite{PhysRevA.84.022326}, and some related variants~\cite{PhysRevA.102.043707,xu2019generalizable,PhysRevA.103.042615} offer many possibilities for the solution.
The pertinent discussions belong to quantum optimal control theory~\cite{PhysRevA.68.062308}, which will be the focus of  our next work.

\begin{acknowledgments}
	This research is supported by the
	Shaanxi Natural Science Basic Research Program (Grant No. 2021JQ-008);
	National Natural Science Foundation of China (NSFC) (Grants No. 12074307 and No. 62071363);
	and China Postdoctoral Science Foundation (Grant No. 2020M673366).\\
\end{acknowledgments}

\appendix
\section{Rewriting of weak commutation condition}\label{Sec:AppA}
Here a new SLD operator $\widetilde{L}_\ell=U_{tot}^\dagger L_\ell U_{tot}$ is introduced at the beginning.
Then the Lyapunov equation associated with the encoded state $\hat{\rho}_\mathbf{x}$ is
\begin{eqnarray}
\partial_\ell \hat{\rho}_\mathbf{x}&=&\frac{\hat{\rho}_\mathbf{ x} \hat{L}_\ell +\hat{L}_\ell \hat{\rho}_\mathbf{ x}}{2}\nonumber\\
&=&\frac{\hat{U}_{tot}\hat{\rho}_\text{in}\hat{U}^\dagger_{tot} \hat{L}_\ell +\hat{L}_\ell \hat{U}_{tot}\hat{\rho}_\text{in} \hat{U}_{tot}^\dagger}{2}\nonumber\\
&=&(\partial_\ell \hat{U}_{tot})\hat{\rho}_\text{in} \hat{U}^\dagger_{tot}+\hat{U}_{tot} \hat{\rho}_\text{in} (\partial_\ell \hat{U}^\dagger_{tot})\;,
\end{eqnarray}
which yields
\begin{eqnarray}\label{eq:SLD}
\hat{L}_\ell=2(\partial_\ell \hat{U}_{tot})\hat{U}_{tot}^\dagger\;.
\end{eqnarray}
\\Combining Eq.~(\ref{eq:SLD}) with $\hat{\mathcal{H}}_\ell:=i(\partial_\ell \hat{U}_{tot}^\dagger)\hat{U}_{tot}=-i\hat{U}_{tot}^\dagger (\partial_\ell \hat{U}_{tot})$, we have
\begin{eqnarray}\label{eq:SLDnew}
\widetilde{L}_\ell=2i\hat{\mathcal{H}}_\ell\;.
\end{eqnarray}
The weak commutation condition (i.e., Eq.~(\ref{eq:weak_comm})) can be rewritten as
\begin{eqnarray}
\text{Tr}\left[\left[\hat{L}_\ell,\hat{L}_{\ell'}\right]\hat{\rho}_{\mathbf{x}}\right]&=&\text{Tr}\left[\left[\widetilde{L}_\ell,\widetilde{L}_{\ell'}\right] \hat{\rho}_\text{in}\right]\nonumber\\
&=&-4\text{Tr}\left[\left[\hat{\mathcal{H}}_\ell, \hat{\mathcal{H}}_{\ell'}\right] \hat{\rho}_\text{in}\right]=0\;,
\end{eqnarray}
where the last line employs Eq.~(\ref{eq:SLDnew}). We thus get the form of Eq.~(\ref{eq:wcc}) in the maintext.\\

\section{Compact form of $\hat{\mathcal{H}}_\ell$}\label{Sec:AppH}
Eq.~(\ref{eq:HY}) is deduced by the following procedures.
According to Eq.~(\ref{eq:Hnew}), one can write out
\begin{eqnarray}\label{eq:Ntimes}
\mathbf{X}_{ \times n} \left(\partial_\ell \mathbf{X}\right)\cdot \vec{J} \!=\!
\left\lbrace \begin{matrix}
\partial_\ell \mathbf{X} \cdot \vec{J},&n=0\\\\
\left|\mathbf{X} \right|^n \left|\partial_\ell \mathbf{X} \right| |\sin\alpha_\ell| \left(\vec{v}_{\ell,n}\cdot\vec{J}\right),&n>0
\end{matrix}\right.\;,\nonumber\\
\end{eqnarray}
where $	\alpha_\ell$ is the angle between vectors $\mathbf{X}$ and $\partial_\ell \mathbf{X}$.
$\vec{v}_{\ell,n}$ represents a unit vector along the direction of $\mathbf{X}_{ \times n} \left(\partial_\ell \mathbf{X}\right)$,
its expression for different $n$ satisfies
$\vec{v}_{\ell,n}=\left\lbrace \begin{matrix}
	-i^{n+1}\; \vec{v}_{\ell,1}\;, & \text{n=odd}\\	
	-i^{n}\; \vec{v}_{\ell,2}\;, & \text{n=even}
\end{matrix}\right.$
where $	\vec{v}_{\ell,1}=\frac{\mathbf{X}_\times \partial_\ell \mathbf{X}}{\left| \mathbf{X}\right| \left| \partial_\ell \mathbf{X}\right| \sin \alpha_\ell }$ and $  \vec{v}_{\ell,2}=\frac{\mathbf{X}_{ \times 2}
	\left(\partial_\ell \mathbf{X}\right)}{\left| \mathbf{X}\right|^2 \left| \partial_\ell \mathbf{X}\right| | \sin \alpha_\ell|}$. 
Plugging Eq.~(\ref{eq:Ntimes}) into Eq.~(\ref{eq:Hnew}), we have
\begin{widetext}
\begin{eqnarray}\label{eq:Hformer}
	\hat{\mathcal{H}}_\ell&=&\sum_{n=0}^{\infty} \frac{(-T)^{n+1}}{(n+1)!}\left( \mathbf{X}_{\times n} \left( \partial_\ell \mathbf{X}\right) \right) \cdot \vec{J}\nonumber\\
	&=&(-T) \partial_\ell \mathbf{X}\cdot \vec{J}+\frac{(-T)^2}{2!} |\mathbf{X}||\partial_\ell \mathbf{X}|\sin\alpha_\ell \left(\vec{v}_{\ell,1} \cdot \vec{J}\right)\nonumber\\
	&+&\frac{(-T)^3}{3!} |\mathbf{X}|^2 |\partial_\ell \mathbf{X}| |\sin \alpha_\ell| \left(\vec{v}_{\ell,2} \cdot \vec{J}\right)+\frac{(-T)^4}{4!}  |\mathbf{X}|^3 |\partial_\ell \mathbf{X}| |\sin \alpha_\ell| \left(\vec{v}_{\ell,3} \cdot \vec{J}\right)+\cdots\nonumber\\
	&=&(-T)\partial_\ell \mathbf{X}\cdot \vec{J}+\frac{(-T)^3}{3!} |\mathbf{X}|^2 |\partial_\ell \mathbf{X}| |\sin \alpha_\ell| \left(\vec{v}_{\ell,2} \cdot \vec{J}\right)\nonumber\\
	&-&\frac{(-T)^5}{5!} |\mathbf{X}|^4 |\partial_\ell \mathbf{X}| |\sin\alpha_\ell| \left(\vec{v}_{\ell,2} \cdot \vec{J}\right)+\frac{(-T)^7}{7!} |\mathbf{X}|^6 |\partial_\ell \mathbf{X}| |\sin\alpha_\ell| \left(\vec{v}_{\ell,2} \cdot \vec{J}\right)+\cdots\nonumber\\
	&+&\frac{(-T)^2}{2!} |\mathbf{X}| |\partial_\ell \mathbf{X}| \sin\alpha_\ell \left(\vec{v}_{\ell,1} \cdot \vec{J}\right)-\frac{(-T)^4}{4!} |\mathbf{X}|^3 |\partial_\ell \mathbf{X}| |\sin\alpha_\ell| \left(\vec{v}_{\ell,1} \cdot \vec{J}\right)\nonumber\\
	&+&\frac{(-T)^6}{6!} |\mathbf{X}|^5 |\partial_\ell \mathbf{X}| |\sin\alpha_\ell| \left(\vec{v}_{\ell,1} \cdot \vec{J}\right)-\frac{(-T)^8}{8!} |\mathbf{X}|^7 |\partial_\ell \mathbf{X}| |\sin\alpha_\ell| \left(\vec{v}_{\ell,1} \cdot \vec{J}\right)+\cdots\nonumber\\
	&=&(-T)\partial_\ell \mathbf{X}\cdot \vec{J}+\frac{|\partial_\ell \mathbf{X}| |\sin\alpha_\ell|}{|\mathbf{X}|}\left[ 
	-\frac{(T|\mathbf{X}|)^3}{3!}+\frac{(T|\mathbf{X}|)^5}{5!}-\frac{(T|\mathbf{X}|)^7}{7!}+\cdots
	\right] \left(\vec{v}_{\ell,2} \cdot \vec{J}\right)\nonumber\\
	&+&\frac{|\partial_\ell \mathbf{X}| |\sin\alpha_\ell|}{|\mathbf{X}|}\left[\frac{(T|\mathbf{X}|)^2}{2!}-\frac{(T|\mathbf{X}|)^4}{4!}+\frac{(T|\mathbf{X}|)^6}{6!}-\cdots
	\right] \left(\vec{v}_{\ell,1} \cdot \vec{J}\right)\nonumber\\
	&=&(-T)\partial_\ell \mathbf{X}\cdot \vec{J}+\frac{|\partial_\ell \mathbf{X}||\sin \alpha_\ell|}{|\mathbf{X}|}
	\Big\{
	[ \sin(T|\mathbf{X}|)-T|\mathbf{X}|] \vec{v}_{\ell,2} +\left[1-\cos(T|\mathbf{X}|) \right]  \vec{v}_{\ell,1}
	\Big\} \cdot \vec{J}\nonumber\\
	&=&|\mathbf{Y}_\ell| \; \vec{e}_\ell \cdot \vec{J}\;,
\end{eqnarray}
where $\vec{e}_\ell$ is a unit vector 
\begin{eqnarray}
\vec{e}_\ell=\frac{1}{|\mathbf{Y}_\ell|} \left\lbrace 
-T ( \partial_\ell \mathbf{X})+\frac{|\partial_\ell \mathbf{X}||\sin \alpha|}{|\mathbf{X}|}
\Big\{[ \sin(T|\mathbf{X}|)-T|\mathbf{X}|] \vec{v}_{\ell,2} +\left[1-\cos(T|\mathbf{X}|) \right]  \vec{v}_{\ell,1}
\Big\} \right\rbrace\;,
\end{eqnarray}
with
\begin{eqnarray}\label{Y}
|\mathbf{Y}_\ell|=\sqrt{T^2 |\partial_\ell \mathbf{X} |^2 (\cos \alpha)^2+\frac{4|\partial_\ell \mathbf{X}|^2 |\sin \alpha|^2}{|\mathbf{X}|^2} \sin^2 \left(\frac{T|\mathbf{X}|}{2} \right)}\;.
\end{eqnarray}
\end{widetext}
To obtain Eq.~(\ref{Y}) we use the following hints
\begin{eqnarray}
	\frac{|\mathbf{X}_{\times2}(\partial_\ell \mathbf{X})|^2}{|\mathbf{X}|^2}=|\mathbf{X}_\times (\partial_\ell \mathbf{X})|^2\;, \\
	\frac{\left(\mathbf{X}_{\times2}(\partial_\ell \mathbf{X})\right) \left(\mathbf{X}_\times \partial_\ell \mathbf{X}\right)}{|\mathbf{X}|}=0\;,\\ 
	\frac{\partial_\ell \mathbf{X}  (\mathbf{X}_{\times2}(\partial_\ell \mathbf{X}))}{|\mathbf{X}|}=\frac{-|\mathbf{X}\times \partial_\ell \mathbf{X}|^2}{|\mathbf{X}|}\;.
\end{eqnarray}

\section{Attainability of the simultaneous optimal estimation in the scheme without controls}\label{Sec:Appwc}
This section mainly elucidates that for the scheme without controls, using a maximally entangled two-qubit state can unconditionally saturate the QCRB of each unknown parameter, but a two-dimensional pure state cannot.
This conclusion is similar to the one proposed in Sec.~\ref{Sec:Ancillary} for the control-enhanced scheme.
Within the following discussions, the weak commutation conditions (i.e., Eq.~(\ref{eq:wcc})) with respect to two different kinds of probe states are analyzed respectively.
\subsection{Two-dimensional pure state} 
From Eq.~(\ref{eq:HY}) we know 
\begin{eqnarray}\label{eq:con}
	\left[\hat{\mathcal{H}}_\ell, \hat{\mathcal{H}}_{\ell'}\right]=i|\mathbf{Y}_\ell| |\mathbf{Y}_{\ell'}| (\vec{e}_\ell \times \vec{e}_{\ell'}) \cdot \vec{J}\;,
\end{eqnarray}
which is inserted into Eq.~(\ref{eq:wcc}), we can get
\begin{eqnarray}\label{eq:weakcondition}
	\text{Tr}\left[\left[\hat{\mathcal{H}}_\ell, \hat{\mathcal{H}}_{\ell'}\right] \hat{\rho}_\text{in}\right]\!=\!\frac{i|\mathbf{Y}_\ell | |\mathbf{Y}_{\ell'} |}{2} (\vec{e}_\ell \times \vec{e}_{\ell'}) \cdot \vec{r}_\text{in}, 
\end{eqnarray}
for $\ell \neq {\ell'}$.
Eq.~(\ref{eq:weakcondition}) means that the QCRBs of $x_\ell$ and $x_{\ell'}$ cannot be simultaneously reached unless the condition $(\vec{e}_\ell \times \vec{e}_{\ell'}) \cdot \vec{r}_\text{in}=0$ is satisfied. 
As a matter of fact, this condition can be removed if using a maximally entangled two-qubit state as the probe state and introducing an ancillary channel in the scheme,  as shown in Fig.~\ref{Fig_scheme} (b). 
\subsection{Maximally entangled two-qubit state} 
Combining  Eq.~(\ref{eq:new_totalU}) with Eq.~(\ref{eq:partH1}) and repeating the computation procedures presented in Sec.~\ref{Sec:withoutC}, Eq.~(\ref{eq:HY}) can be replaced with
\begin{eqnarray}\label{eq:HoutA}
	\hat{\mathcal{H}}_{\ell,A}=|\mathbf{Y}_\ell| \;  \vec{e}_\ell \cdot \vec{J} \otimes \hat{I}_A\;,
\end{eqnarray}
where $\hat{I}_A$ represents the introduced ancillary channel. 
Eq.~(\ref{eq:HoutA}) indicates
\begin{eqnarray}\label{eq:wc}
	[\hat{\mathcal{H}}_{\ell,A},\hat{\mathcal{H}}_{{\ell'},A}]&=&|\mathbf{Y}_\ell| |\mathbf{Y}_{\ell'}|[\vec{e}_\ell \cdot \vec{J},\; \vec{e}_{\ell'} \cdot \vec{J}] \otimes \hat{I}_A\nonumber\\
	&=&i |\mathbf{Y}_\ell | |\mathbf{Y}_{\ell'} | (\vec{e}_\ell \times \vec{e}_{\ell'}) \cdot \vec{J} \otimes \hat{I}_A\;.
\end{eqnarray}
Then inserting a pure state $|\psi_{SA}\rangle$ into Eq.~(\ref{eq:wcc}) we have
\begin{eqnarray}\label{eq:wek}
	&&\text{Tr}\left[\hat{\rho}_\text{in} [\hat{\mathcal{H}}_{\ell,A},\hat{\mathcal{H}}_{\ell',A}]  \right]\nonumber\\
	&=& \text{Tr}\left[\hat{\rho}_\text{in} \left(i |\mathbf{Y}_\ell| |\mathbf{Y}_{\ell'}| (\vec{e}_\ell \times \vec{e}_{\ell'})
	\cdot \vec{J} \otimes \hat{I}_A\right)\right]\nonumber\\
	&=& \text{Tr}\left[\hat{\rho}_S  \left(i |\mathbf{Y}_\ell| |\mathbf{Y}_{\ell'}| (\vec{e}_\ell \times \vec{e}_{\ell'}) \cdot \vec{J}\right)\right]\;,
\end{eqnarray}
where $\hat{\rho}_S$ is the reduced density operator of $\hat{\rho}_\text{in}$ after tracing out the ancillary part, i.e., $\hat{\rho}_S=\text{Tr}_A[\hat{\rho}_\text{in}]=\text{Tr}_A\left[|\psi_{SA}\rangle \langle \psi_{SA}|\right]$.
Diagonalizing the matrix $i |\mathbf{Y}_\ell| |\mathbf{Y}_{\ell'}| (\vec{e}_\ell \times \vec{e}_{\ell'}) \cdot \vec{J}$, we have
\begin{eqnarray}\label{eq:dia}
	&&i |\mathbf{Y}_\ell| |\mathbf{Y}_{\ell'}| (\vec{e}_\ell \times \vec{e}_{\ell'}) \cdot \vec{J}\nonumber\\
	&=&\hat{V} {\hspace{-0.5mm}
		\left( \begin{matrix}
			+c  |\mathbf{Y}_\ell| |\mathbf{Y}_{\ell'}| |\vec{e}_\ell \times \vec{e}_{\ell'}|&0\\\\
			0&-c  |\mathbf{Y}_\ell| |\mathbf{Y}_{\ell'}| |\vec{e}_\ell \times \vec{e}_{\ell'}|
		\end{matrix}\right)}\hat{V}^\dagger,\nonumber\\
\end{eqnarray}
where $\hat{V}$ is a unitary matrix and $+c$ ($-c$) is the maximal (minimal) eigenvalue of $\hat{j}_m$ for $m =1,2,3$. 
Plugging Eq.~(\ref{eq:dia}) into Eq.~(\ref{eq:wek}) and using the new quantum state $\hat{\rho}^{\prime\prime}=\hat{V}^\dagger \hat{\rho}_S \hat{V}$ with the diagonal elements ${\rho}_{11}^{\prime\prime}$, ${\rho}_{22}^{\prime\prime}$,  we can get 
\begin{eqnarray}\label{eq:newwcc}
	&&\text{Tr} \left[\hat{\rho}_\text{in}  [\hat{\mathcal{H}}_{\ell,A},\hat{\mathcal{H}}_{\ell',A}]  \right]\nonumber\\
	&=&\text{Tr}
	\!	\!\left[\hat{\rho}^{\prime\prime} \!\! \left( \begin{matrix}
		+c  |\mathbf{Y}_\ell| |\mathbf{Y}_{\ell'}| |\vec{e}_\ell \times \vec{e}_{\ell'}|   & 0\\\\
		0  &  -c  |\mathbf{Y}_\ell| |\mathbf{Y}_{\ell'}| |\vec{e}_\ell \times \vec{e}_{\ell'}|
	\end{matrix}\right)\!\!  \right]  \nonumber\\
	&=&({\rho}^{\prime\prime}_{11}-{\rho}^{\prime\prime}_{22}) c |\mathbf{Y}_\ell| |\mathbf{Y}_{\ell'}| |\vec{e}_\ell \times \vec{e}_{\ell'}|\;,
\end{eqnarray}
The result of Eq.~(\ref{eq:newwcc}) equals to zero if $\hat{\rho}{''}=\hat{I}/2$. It implies $\hat{\rho}_S=\hat{I}/2$ so that $\hat{\rho}_\text{in}$ is a maximally entangled two-qubit state.
Compared with Eq.~(\ref{eq:weakcondition}) confined by the condition $(\vec{e}_m \times \vec{e}_n) \cdot \vec{r}_\text{in}=0$, Eq.~(\ref{eq:newwcc}) can unconditionally equal to zero if using a maximally entangled two-qubit state and an ancillary channel.

\section{Obtainable QFI maximum}\label{Sec:Appsame}
This section clarifies that taking a two-dimensional pure state or a maximally entangled two-qubit state as the probe state can offer the same QFI maximum.
But the advantage of using the maximally entangled two-qubit state is that this QFI maximum can be unconditionally reached, by contrast, it is restricted if using a two-dimensional pure state.
This clarification is respectively proved in the scheme without and with controls.
\subsection{Without controls}
Taking a maximally entangled two-qubit state $|\psi_{SA}\rangle$ as the probe state, i.e., $\hat{\rho}_\text{in}=|\psi_{SA}\rangle \langle \psi_{SA}|$, according to Eq.~(\ref{eq:HoutA}) we have 
\begin{eqnarray}
	\text{Tr}\left[\hat{\rho}_\text{in} \hat{\mathcal{H}}_{\ell,A}^2 \right]
	&=&|\mathbf{Y}_\ell|^2 \!\text{Tr}\! \left[\hat{\rho}_\text{in} \left(\vec{e}_\ell \cdot \vec{J}\right)^2 \otimes \hat{I}_A
	 \right]=\frac{|\mathbf{Y}_\ell|^2}{4}\;,\label{eq:ecc}\nonumber\\\\
	\text{Tr}\left[ \hat{\rho}_\text{in} \hat{\mathcal{H}}_{\ell,A} \right]
	&=&\langle \psi_{SA}| (|\mathbf{Y}_\ell | \; \vec{e}_\ell \cdot \vec{J}\otimes \hat{I}_A)| \psi_{SA}\rangle\nonumber\\
	&=& \text{Tr}\left[\hat{\rho}_S \left( |\mathbf{Y}_\ell | \; \vec{e}_\ell \cdot \vec{J}\right)\right]\label{eq:trace1}\;,
\end{eqnarray}
where $\hat{\rho}_S=\text{Tr}_A[|\psi_{SA}\rangle \langle \psi_{SA}|]$.
Then the matrix $|\mathbf{Y}_\ell | \; \vec{e}_\ell \cdot \vec{J}$ is diagonalized as 
\begin{eqnarray}\label{eq:diagonalize}
	|\mathbf{Y}_\ell | \; \vec{e}_\ell \cdot \vec{J}=\hat{\mathcal{Y}}
	\left( \begin{matrix}
		+c |\mathbf{Y}_\ell |&0\\
		0&-c |\mathbf{Y}_\ell |
	\end{matrix}\right) \hat{\mathcal{Y}}^\dagger\;,
\end{eqnarray}
where $\hat{\mathcal{Y}}$ is a unitary matrix, 
$+c$ ($-c$) is the maximal (minimal) eigenvalue of $\hat{j}_m$ for $m =1,2,3$. 
Inserting Eq.~(\ref{eq:diagonalize}) into Eq.~(\ref{eq:trace1}) and introducing the new quantum state $\hat{\varrho}=\hat{\mathcal{Y}}^\dagger \hat{\rho}_S \hat{\mathcal{Y}}$ with the diagonal elements ${\varrho}_{11}$, ${\varrho}_{22}$, we have
\begin{eqnarray}\label{eq:tracenew}
	\text{Tr}\left[ \hat{\rho}_\text{in} \hat{\mathcal{H}}_{\ell,A} \right]&=&\text{Tr}\left[\hat{\varrho}
	 \left( \begin{matrix}
		+c |\mathbf{Y}_\ell |&0\\
		0&-c |\mathbf{Y}_\ell |
	\end{matrix}\right) 
	\right]\nonumber\\
	&=&({\varrho}_{11}-{\varrho}_{22}) c  |\mathbf{Y}_\ell |\;,
\end{eqnarray}
which equals to zero since $\hat{\rho}_S=\hat{\varrho}=\hat{I}/2$.

After interacting with the entire dynamics $\hat{U}_{tot}$, we get $\hat{\rho}_\mathbf{x}=|\psi_\mathbf{x}\rangle \langle \psi_\mathbf{x}|=\hat{U}_{tot}|\psi_{SA}\rangle \langle \psi_{SA}| \hat{U}^\dagger_{tot}$.
The QFI of the parameter $x_\ell$ is~\cite{Review}
\begin{eqnarray}\label{eq:pure}
\mathcal{I}_{\ell\ell}&=&4\left(\langle \partial_\ell \psi_\mathbf{x}|\partial_\ell \psi_\mathbf{x}\rangle-\left|\langle\partial_\ell \psi_\mathbf{x}|\psi_\mathbf{x}\rangle\right|^2\right)\nonumber\\
&=&4\left(\langle\psi_{SA}|\hat{\mathcal{H}}_\ell \hat{\mathcal{H}}^\dagger_\ell |\psi_{SA}\rangle -|\langle \psi_{SA}|(-i\hat{\mathcal{H}}_\ell)|\psi_{SA}\rangle |^2\right)\nonumber\\
&=&4\left( \text{Tr}\left[ \hat{\mathcal{H}}_\ell^2 \; \hat{\rho}_\text{in}\right]-\text{Tr}^2 \left[\hat{\mathcal{H}}_\ell \; \hat{\rho}_\text{in} \right]\right)\;,
\end{eqnarray} 
where the second line uses the definition of Eq.~(\ref{eq:def}), the result of the last line is consistent with Eq.~(\ref{eq:QFIM_pure}).
Inserting Eqs.~(\ref{eq:ecc}) and (\ref{eq:tracenew}) into Eq.~(\ref{eq:pure}), we finally get
\begin{eqnarray}\label{eq:QFI_maxA}
	\mathcal{I}_{\ell\ell}=|\mathbf{Y}_\ell |^2\;,
\end{eqnarray}
which equals to Eq.~(\ref{eq:QFI_max}), more interestingly, it has no requirement for $\vec{e}_\ell \cdot \vec{r}_\text{in}=0$.
\subsection{With controls}
In the control-aided case, the counterpart of Eq.~(\ref{eq:HoutA}) is $\widetilde{\mathcal{H}}_{\ell,A}=-T \left(\partial_\ell \mathbf{X} \cdot \vec{J}\right)\otimes \hat{I}_A$ (see Eq.~(\ref{eq:H1})). 
With the maximally entangled two-qubit state, we obtain
\begin{eqnarray}
	\text{Tr}\left[\hat{\rho}_\text{in} \widetilde{\mathcal{H}}_{\ell,A}^2 \right]=\frac{T^2 (\partial_\ell \mathbf{X})^2}{4}, \quad 
	\text{Tr}\left[\hat{\rho}_\text{in} \widetilde{\mathcal{H}}_{\ell,A} \right]=0.\label{eq:new2}
\end{eqnarray}
\\The corresponding QFI can be derived by inserting Eq.~(\ref{eq:new2}) into Eq.~(\ref{eq:pure}), i.e.,  
\begin{eqnarray}\label{eq:QFI_new}
	\widetilde{\mathcal{I}}_{\ell \ell}=T^2|\partial_\ell \mathbf{X}|^2\;.
\end{eqnarray}
Likewise, Eq.~(\ref{eq:QFI_new}) equals to Eq.~(\ref{eq:QFImWC}) and has no requirement for $\partial_\ell \mathbf{X} \cdot \vec{r}_{\text{in}}=0$.

\section{Off-diagonal elements of the QFIM}\label{Sec:APPoff}
For the no-control scheme with the assumption of $\vec{e}_B \cdot \vec{r}_{\text{in}}=\vec{e}_\theta \cdot \vec{r}_{\text{in}}= \vec{e}_\phi \cdot \vec{r}_{\text{in}}=0$.  
Combining $\hat{\mathcal{H}}_B$ (Eq.~(\ref{eq:HB})), $\hat{\mathcal{H}}_\theta$ (Eq.~(\ref{eq:Htheta})) and $\hat{\mathcal{H}}_\phi$  (Eq.~(\ref{eq:H_phi})) with  Eq.~(\ref{eq:cov}) we get
\begin{widetext}
	\begin{eqnarray}
		\text{Cov}_\text{in} (\hat{\mathcal{H}}_B,\hat{\mathcal{H}}_\theta
		)&=&\frac{1}{2}\text{Tr}\left[\{\hat{\mathcal{H}}_B, \hat{\mathcal{H}}_\theta \} \hat{\rho}_{\text{in}} \right]-\text{Tr}\left[\hat{\mathcal{H}}_B \hat{\rho}_{\text{in}}  \hat{\mathcal{H}}_\theta \hat{\rho}_\text{in}  \right]\nonumber\\
		&=&T\sin(BT)\left(\vec{e}_B \cdot \vec{e}_\theta \right)-\frac{T\sin(BT)}{4}\text{Tr}\left[ (\vec{e}_B \cdot \vec{e}_\theta)\hat{I}-(\vec{e}_B \times \vec{r}_{\text{in}})\cdot (\vec{e}_\theta \times \vec{r}_\text{in}) \hat{I}
		\right]=0\;,\label{cov1}\\
		\text{Cov}_\text{in} (\hat{\mathcal{H}}_B,\hat{\mathcal{H}}_\phi
		)&=&\frac{1}{2}\text{Tr}\left[\{\hat{\mathcal{H}}_B, \hat{\mathcal{H}}_\phi \} \hat{\rho}_\text{in} \right]-\text{Tr}\left[\hat{\mathcal{H}}_B \hat{\rho}_\text{in}  \hat{\mathcal{H}}_\phi \hat{\rho}_\text{in}  \right]\nonumber\\
		&=&T\sin(BT)\sin\theta(\vec{e}_B \cdot \vec{e}_\phi)-{T\sin(BT)\sin\theta}\; \text{Tr} \Big[\left(\vec{e}_B \cdot \vec{J}\right)\left(\vec{e}_\phi \cdot \vec{J}\right)\nonumber\\
		&+&\left(\vec{e}_B \cdot \vec{J}\right)\left((\vec{e}_\phi \times \vec{r}_\text{in})\cdot \vec{J}\right)\!+\!\left((\vec{e}_B \times \vec{r}_\text{in}) \cdot \vec{J}\right)\left(\vec{e}_\phi \cdot \vec{J}\right)+\left((\vec{e}_B \times \vec{r}_\text{in})\cdot \vec{J}\right)\left((\vec{e}_\phi \times \vec{r}_\text{in}) \cdot \vec{J}\right)\Big]\!=\!0, \label{cov2}\\
		\text{Cov}_\text{in} (\hat{\mathcal{H}}_\theta,\hat{\mathcal{H}}_\phi
		)&=&\frac{1}{2}\text{Tr}\left[\{\hat{\mathcal{H}}_\theta, \hat{\mathcal{H}}_\phi \} \hat{\rho}_\text{in} \right]-\text{Tr}\left[\hat{\mathcal{H}}_\theta \hat{\rho}_\text{in}  \hat{\mathcal{H}}_\phi \hat{\rho}_\text{in}  \right]\nonumber\\
		&=&\sin^2(BT)\sin\theta (\vec{e}_\theta \cdot \vec{e}_\phi)
		-{\sin^2 (BT)\sin \theta}\; \text{Tr}\Big[\left(\vec{e}_\theta \cdot \vec{J}\right)\left(\vec{e}_\phi \cdot \vec{J}\right) \nonumber\\
		& +& i \left(\vec{e}_\theta \cdot \vec{J}\right)\left((\vec{e}_\phi \times \vec{r}_\text{in})\cdot \vec{J}\right)+\! i \! \left((\vec{e}_\theta \times \vec{r}_\text{in})\cdot \vec{J}\right)\left(\vec{e}_\phi \cdot \vec{J}\right)\!-\!\left((\vec{e}_\theta \times \vec{r}_\text{in}) \cdot \vec{J}\right)\left((\vec{e}_\phi \times \vec{r}_\text{in})\cdot \vec{J}\right)\Big]\!=\!0.\label{cov3}
	\end{eqnarray}
Inserting Eqs.~(\ref{cov1})-(\ref{cov3}) into Eq.~(\ref{eq:cov_pure}),  the off-diagonal entries of the QFIM thus are given by
\begin{eqnarray}
	\mathcal{F}_{B\theta}=4 \text{Cov}_\text{in} (\hat{\mathcal{H}}_B,\hat{\mathcal{H}}_\theta)=0\;,\quad
	\mathcal{F}_{B\phi}=4 \text{Cov}_\text{in} (\hat{\mathcal{H}}_B,\hat{\mathcal{H}}_\phi)=0\;,\quad
	\mathcal{F}_{\theta\phi}=4 \text{Cov}_\text{in} (\hat{\mathcal{H}}_\theta,\hat{\mathcal{H}}_\phi)=0\;.
\end{eqnarray}

Besides, for the control-aided scheme with the assumption of $\vec{n}_0 \cdot \vec{r}_{\text{in}}=\vec{n}^{'}_0 \cdot \vec{r}_{\text{in}}=\vec{n}^{''}_0 \cdot \vec{r}_{\text{in}}=0$.
Combining $\widetilde{\mathcal{H}}_B$ (Eq.~(\ref{eq:HBnew})), $\widetilde{\mathcal{H}}_\theta$ (Eq.~(\ref{eq:Hthetanew})) and $\widetilde{\mathcal{H}}_\phi$  (Eq.~(\ref{eq:Hphinew})) with Eq.~(\ref{eq:cov}) we have
	\begin{eqnarray}
		{\text{Cov}}_\text{in} (\widetilde{\mathcal{H}}_B,\widetilde{\mathcal{H}}_\theta
		)&=&\frac{1}{2}\text{Tr}\left[\{\widetilde{\mathcal{H}}_B, \widetilde{\mathcal{H}}_\theta \} \hat{\rho}_{\text{in}} \right]-\text{Tr}\left[\widetilde{\mathcal{H}}_B \hat{\rho}_{\text{in}}  \widetilde{\mathcal{H}}_\theta \hat{\rho}_\text{in}  \right]\nonumber\\
		&=&BT^2(\vec{n}_0 \cdot \vec{n}^{'}_0)-\frac{BT^2}{4}\text{Tr}\left[(\vec{n}_0 \cdot \vec{n}^{'}_0 ) \hat{I}-(\vec{n}_0 \times \vec{r}_\text{in})\cdot (\vec{n}^{'}_0 \times \vec{r}_\text{in}) \hat{I}
		\right]=0\;,\label{cov1new}\\
		{\text{Cov}}_\text{in} (\widetilde{\mathcal{H}}_B,\widetilde{\mathcal{H}}_\phi
		)&=&\frac{1}{2}\text{Tr}\left[\{\widetilde{\mathcal{H}}_B, \widetilde{\mathcal{H}}_\phi \} \hat{\rho}_\text{in} \right]-\text{Tr}\left[\widetilde{\mathcal{H}}_B \hat{\rho}_\text{in}  \widetilde{\mathcal{H}}_\phi \hat{\rho}_\text{in}  \right]\nonumber\\
		&=&BT^2(\vec{n}_0 \cdot \vec{n}^{''}_0)-\frac{BT^2}{4}\text{Tr}\left[(\vec{n}_0 \cdot \vec{n}^{''}_0 ) \hat{I}-(\vec{n}_0 \times \vec{r}_\text{in})\cdot (\vec{n}^{''}_0 \times \vec{r}_\text{in}) \hat{I}
		\right]=0\;,\label{cov2new}\\
		{\text{Cov}}_\text{in} (\widetilde{\mathcal{H}}_\theta,\widetilde{\mathcal{H}}_\phi
		)&=&\frac{1}{2}\text{Tr}\left[\{\widetilde{\mathcal{H}}_\theta, \widetilde{\mathcal{H}}_\phi \} \hat{\rho}_{\text{in}} \right]-\text{Tr}\left[\widetilde{\mathcal{H}}_\theta \hat{\rho}_{\text{in}}  \widetilde{\mathcal{H}}_\phi \hat{\rho}_\text{in}  \right]\nonumber\\
		&=&(BT)^2 (\vec{n}^{'}_0 \cdot \vec{n}^{''}_0)-\frac{(BT)^2}{4} \text{Tr}\left[(\vec{n}^{'}_0 \cdot \vec{n}^{''}_0) \hat{I}-(\vec{n}^{'}_0 \times \vec{r}_\text{in})\cdot (\vec{n}^{''}_0 \times \vec{r}_\text{in})\hat{I}\right]=0\;.\label{cov3new}
	\end{eqnarray}
Substituting Eqs.~(\ref{cov1new})-(\ref{cov3new}) into Eq.~(\ref{eq:cov_pure}),  the off-diagonal entries of the QFIM read
\begin{eqnarray}
\widetilde{\mathcal{F}}_{B\theta}=4 {\text{Cov}}_\text{in} (\widetilde{\mathcal{H}}_B,\widetilde{\mathcal{H}}_\theta)=0\;,\quad
\widetilde{\mathcal{F}}_{B\phi}=4 {\text{Cov}}_\text{in} (\widetilde{\mathcal{H}}_B,\widetilde{\mathcal{H}}_\phi)=0\;,\quad
\widetilde{\mathcal{F}}_{\theta\phi}=4 {\text{Cov}}_\text{in} (\widetilde{\mathcal{H}}_\theta,\widetilde{\mathcal{H}}_\phi)=0\;.
\end{eqnarray}

\section{Summary of the relevant QFI results}\label{App:tables}
Here we summarize the relevant QFI results in two tables to indicate the effectiveness of quantum control in promoting  the estimation precision of multiple parameters. 
Compared with the last column of Table.\ref{table:without}, the QFI results at the last column of Table.\ref{table:with} implies that the estimation precision with respect to  $\theta$ and $\phi$  can be promoted to the Heisenberg scaling $1/T$.
\begin{table*}[h]
\centering
\caption{\label{table:without}The QFI results without any control operation}
\begin{ruledtabular}
\begin{tabular}{ccc}
{Scheme}&\multicolumn{1}{c}{Fig.~\ref{Fig_scheme}(a) } & Fig.~\ref{Fig_scheme}(b)\\
{Probe state}&Two-dimensional  pure  state&Maximally entangled two-qubit state\\
\hline
$\hat{H}(\mathbf{x})=\mathbf{X}\cdot \vec{J}$&$	\mathcal{F}_{\ell\ell}=|\mathbf{Y}|^2 \left(1-(\vec{e}_\ell \cdot \vec{r}_\text{in})^2\right)$&$	\mathcal{F}^{\text {max}}_{\ell \ell}=|\mathbf{Y}|^2$\\
\hline
&$\mathcal{F}_B=4T^2 |\vec{r}_{\text{in}}|^2  \left(1-(\vec{e}_B \cdot \vec{r}_1)^2\right)$&$	\mathcal{F}_B^{\text{max}}=4T^2$\\
$\hat{H}(\mathbf{x})=2B\vec{n}_0 \cdot \vec{J}$&$\mathcal{F}_\theta=4\sin^2(BT)\left(1-(\vec{e}_\theta \cdot \vec{r}_{\text{in}})^2\right)$&$\mathcal{F}^{\text{max}}_\theta=4\sin^2 (BT)$\\
&$\mathcal{F}_\phi=4 \sin^2 \theta \sin^2 (BT) |\vec{r}_{\text{in}}|^2 \left(1-(\vec{e}_\phi \cdot \vec{r}_1)^2\right)$&$\mathcal{F}^{\text{max}}_\phi=4\sin^2 \theta\sin^2 (BT)$\\
\end{tabular}
\end{ruledtabular}
\end{table*}
\begin{table*}[h]
\centering
\caption{\label{table:with}The control-enhanced QFI results}
\begin{ruledtabular}
\begin{tabular}{ccc}
{Scheme}&\multicolumn{1}{c}{Fig.~\ref{Fig_scheme}(a) } & Fig.~\ref{Fig_scheme}(b)\\
{Probe state}&Two-dimensional  pure state&Maximally entangled two-qubit state\\
\hline
$\hat{H}(\mathbf{x})=\mathbf{X}\cdot \vec{J}$&$\widetilde{\mathcal{F}}_{\ell \ell}=T^2 \left(|\partial_\ell  \mathbf{X}|^2-(\partial_\ell \mathbf{X} \cdot \vec{r}_\text{in})^2 \right)$&$\widetilde{\mathcal{F}}^{\text {max}}_{\ell \ell}=T^2|\partial_\ell \mathbf{X}|^2$\\
\hline
&$\widetilde{\mathcal{{F}}}_B=4|\vec{r}_{\text{in}}|^2 T^2\left(1-(\vec{n}_0 \cdot \vec{r}_1)^2\right)$&$\widetilde{\mathcal{F}}_B^{\text{max}}=4T^2$\\
$\hat{H}(\mathbf{x})=2B\vec{n}_0 \cdot \vec{J}$&
$\widetilde{\mathcal{F}}_\theta=4(BT)^2 \left(1-(\vec{n}^{'}_0 \cdot \vec{r}_\text{in})^2\right)$&$	\widetilde{\mathcal{F}}_{\theta}^\text{max}=4(BT)^2$\\
&$\widetilde{\mathcal{F}}_\phi=4 (BT)^2|\vec{r}_\text{in}|^2 \left(\sin^2 \theta-(\vec{n}^{''}_0 \cdot \vec{r}_1)^2\right)$&$\widetilde{\mathcal{F}}^{\text {max}}_\phi=4(BT)^2 \sin^2 \theta$\\
\end{tabular}
\end{ruledtabular}
\end{table*}
\end{widetext}


\begin{thebibliography}{100}
	\bibitem{Multi_1} M. Szczykulska, T.  Baumgratz, and A.  Datta, Multi-parameter quantum metrology, Adv. Phys.: X {\bf 1}, 621 (2016).
	
	\bibitem{Multi_2} L. Pezz\`e,  M. A. Ciampini,  N. Spagnolo,  P. C. Humphreys, A.  Datta,  I. A. Walmsley, M.  Barbieri,  F. Sciarrino, and A. Smerzi, Optimal Measurements for Simultaneous Quantum Estimation of Multiple Phases, Phys. Rev. Lett. {\bf 119}, 130504 (2017).
	
	\bibitem{Multi_3} E. Polino,  M. Valeri,  N.  Spagnolo, and F. Sciarrino, Photonic quantum metrology, AVS Quantum Sci. {\bf 2}, 024703 (2020).
	
	\bibitem{Magneticfile0} I. Apellaniz,  I. Urizar-Lanz,  Z. Zimbor\'as, P.  Hyllus, and G. T\'oth, Precision bounds for gradient magnetometry with atomic ensembles, Phys. Rev. A {\bf 97}, 053603 (2018).
	
	\bibitem{PhysRevA.95.012326} P. Kok,  J.  Dunningham, and J. F. Ralph, Role of entanglement in calibrating optical quantum gyroscopes, Phys. Rev. A {\bf 95}, 012326 (2017).
	
	\bibitem{PhysRevApplied.14.034065} 	M. R. Grace, C. N. Gagatsos,  Q.  Zhuang, and S. Guha, Quantum-Enhanced Fiber-Optic Gyroscopes Using Quadrature Squeezing and Continuous-Variable Entanglement, Phys. Rev. Applied {\bf 14}, 034065 (2020).
	
	
	\bibitem{PhysRevLett.120.080501} T. J.  Proctor, P. A.  Knott, and J. A. Dunningham, Multiparameter Estimation in Networked Quantum Sensors, Phys. Rev. Lett. {\bf 120}, 080501 (2018).
	
	\bibitem{Pa} P. C. Humphreys, M. Barbieri,  A.  Datta, and I. A.  Walmsley, Quantum Enhanced Multiple Phase Estimation, Phys. Rev. Lett. {\bf 111}, 070403 (2013).
	
	
	\bibitem{PhysRevA.interferometer}  N. Fabre and S. Felicetti, Parameter estimation of time and frequency shifts with generalized Hong-Ou-Mandel interferometry, Phys. Rev. A {\bf 104}, 022208 (2021).

	\bibitem{PhysRevX.6.031033} M. Tsang,  R. Nair, and X.-M. Lu, Quantum Theory of Superresolution for Two Incoherent Optical Point Sources, Phys. Rev. X {\bf 6}, 031033 (2016).
	
	\bibitem{PhysRevLett.122.140505} C. Napoli,  S.  Piano, R.  Leach, G.  Adesso, and T. Tufarelli, Towards Superresolution Surface Metrology: Quantum Estimation of Angular and Axial Separations, Phys. Rev. Lett. {\bf 122}, 140505 (2019).


	\bibitem{PRXQuantum.2.010301} V. Ansari, B. Brecht, J. Gil-Lopez, J. M. Donohue, J. \ifmmode \check{R}\else \v{R}\fi{}eh\'a\ifmmode \check{c}\else \v{c}\fi{}ek, Z. Hradil,  L. L. S\'anchez-Soto, and C. Silberhorn, Achieving the Ultimate Quantum Timing Resolution, PRX Quantum {\bf 2}, 010301 (2021).
	
	\bibitem{PhysRevLett.112.047601} M. W. Doherty, V. V. Struzhkin, D. A. Simpson, L. P.  McGuinness, Y.  Meng, A.  Stacey, T. J. Karle, R. J. Hemley, N. B.  Manson, L. C. L.  Hollenberg \textit{et al.}, Electronic Properties and Metrology Applications of the Diamond ${\mathrm{NV}}^{\ensuremath{-}}$ Center under Pressure, Phys. Rev. Lett. {\bf 112}, 047601 (2014).
	
	\bibitem{Pang2014} S. Pang and T. A. Brun, Quantum metrology for a general {H}amiltonian parameter, Phys. Rev. A {\bf 90}, 022117 (2014).
	
	\bibitem{Magneticfile1} T. Baumgratz and A. Datta, Quantum Enhanced Estimation of a Multidimensional Field, Phys. Rev. Lett. {\bf 116}, 030801 (2016).
	
	
	\bibitem{PhysRevA.92.012312} X.-X. Jing,  J.  Liu, H.-N.  Xiong,  and X. Wang, Maximal quantum Fisher information for general su(2) parametrization processes, Phys. Rev. A {\bf 92}, 012312 (2015).
	
	\bibitem{PhysRevLett.125.020501} Z. Hou,  Z. Zhang,  G.-Y. Xiang,  C.-F. Li,  G.-C. Guo, H. Chen, L. Liu, and H. Yuan, Minimal Tradeoff and Ultimate Precision Limit of Multiparameter Quantum Magnetometry under the Parallel Scheme, Phys. Rev. Lett. {\bf 125}, 020501 (2020).
	
	
	\bibitem{Se} H. Yuan, Sequential Feedback Scheme Outperforms the Parallel Scheme for Hamiltonian Parameter Estimation, Phys. Rev. Lett. {\bf 117}, 160801 (2016).
	
	\bibitem{Multi_Sequential} Z. Hou, J.-F. Tang, H.  Chen, H. Yuan, G.-Y. Xiang, C.-F.  Li, and G.-C. Guo, Zero--trade-off multiparameter quantum estimation via simultaneously saturating multiple Heisenberg uncertainty relations, Sci. Adv. {\bf 7}, eabd2986 (2021). 
	
	\bibitem{Mu_satu3} S. Ragy,  M.  Jarzyna, and  R. Demkowicz-Dobrza\ifmmode \acute{n}\else \'{n}\fi{}ski, Compatibility in multiparameter quantum metrology, Phys. Rev. A {\bf 94}, 052108 (2016).
	
	\bibitem{Belliardo_2021} F. Belliardo and V. Giovannetti, Incompatibility in quantum parameter estimation, New J.  Phys. {\bf 23}, 063055 (2021).
	
	\bibitem{Kull_2020} I. Kull , P. A. Gu{\'{e}}rin, and F. Verstraete, Uncertainty and trade-offs in quantum multiparameter estimation, J. Phys. A: Math. Theor.  {\bf 53}, 244001 (2020).
	
	\bibitem{albarelli2021probe} F. Albarelli and R. Demkowicz-Dobrza\ifmmode \acute{n}\else \'{n}\fi{}ski, Probe incompatibility in multiparameter noisy quantum channel estimation, arXiv:2104.11264.
	
	\bibitem{pang2017optimal} S. Pang and A. N. Jordan, Optimal adaptive control for quantum metrology with time-dependent Hamiltonians, Nat. Commun. {\bf 8}, 14695 (2017).
	
	\bibitem{Review} J. Liu, H.  Yuan,  X.-M.  Lu, and X. Wang, Quantum Fisher information matrix and multiparameter estimation, J. Phys. A: Math. Theor. {\bf 53}, 023001 (2019).
	
	\bibitem{liu2015quantum} J. Liu, X.-X. Jing, and X. Wang, Quantum metrology with unitary parametrization processes, Sci. Rep. {\bf 5}, 8565  (2015).
	
	\bibitem{QFI} M. G. A. Paris, Quantum estimation for quantum technology, Int. J.  Quantum Inform. {\bf 7}, 125 (2009).
	
	\bibitem{Mu_satu} K. Matsumoto, A new approach to the Cram\'er-Rao-type bound of the pure-state model, J. Phys. A: Math. Gen.  {\bf 35}, 3111 (2002).
	
	\bibitem{Mu_satu2} M. D. Vidrighin, G. Donati, M. G. Genoni,  X.-M.  Jin,  W. S.  Kolthammer,  M. S.  Kim, A. Datta,  M.  Barbieri, and I. A. Walmsley,  Joint estimation of phase and phase diffusion for quantum metrology, Nat. Commun. {\bf 5}, 3532 (2014).
	
	\bibitem{Single_Sequential_Hou} Z. Hou, R.-J. Wang, J.-F. Tang, H. Yuan, G.-Y. Xiang, C.-F.  Li, and G.-C. Guo, Control-Enhanced Sequential Scheme for General Quantum Parameter Estimation at the Heisenberg Limit, Phys. Rev. Lett. {\bf 123}, 040501 (2019).
	
	\bibitem{giovannetti2004quantum} V. Giovannetti, S. Lloyd, and L. Maccone, Quantum-enhanced measurements: beating the standard quantum limit, Science {\bf 306}, 1330 (2004).
	
	\bibitem{QM} V. Giovannetti, S. Lloyd,  and L. Maccone, Advances in quantum metrology, Nat. Photonics {\bf 5}, 222 (2011).
	
	\bibitem{wilcox1967exponential} R. M Wilcox, Exponential operators and parameter differentiation in quantum physic, J.  Math. Phys. {\bf 8}, 962 (1967).
	
	\bibitem{Purity} F. Chapeau-Blondeau, Optimized probing states for qubit phase estimation with general quantum noise, Phys. Rev. A {\bf 91}, 052310 (2015).
	
	
	\bibitem{Paris2021properties} A. Candeloro,  M. G. A. Paris, and M. G Genoni, On the properties of the asymptotic incompatibility measure in multiparameter quantum estimation,  J. Phys. A: Math. Theor. {\bf 54}, 485301 (2021).
	
	\bibitem{PhysRevA.63.042304} A. Fujiwara, Quantum channel identification problem, Phys. Rev. A {\bf 63}, 042304 (2001).
	
	\bibitem{quintino2021deterministic} M. T. Quintino and D. Ebler, Deterministic transformations between unitary operations: Exponential advantage with adaptive quantum circuits and the power of indefinite causality, arXiv:2109.08202.
	
	\bibitem{PhysRevLett.127.200504} J. Bavaresco,  M. Murao, and M. T.  Quintino, Strict Hierarchy between Parallel, Sequential, and Indefinite-Causal-Order Strategies for Channel Discrimination, Phys. Rev. Lett. {\bf 127}, 200504 (2021).
	
	\bibitem{Hourotate} Z. Hou, Y. Jin, H. Chen, J.-F. Tang, C.-J. Huang, H.  Yuan, G.-Y.  Xiang, C.-F.  Li, and G.-C. Guo, ``Super-Heisenberg'' and Heisenberg Scalings Achieved Simultaneously in the Estimation of a Rotating Field, Phys. Rev. Lett. {\bf 126}, 070503 (2021).
	
	\bibitem{Fujiwara_2006} A. Fujiwara, Strong consistency and asymptotic efficiency for adaptive quantum estimation problems, J. Phys. A: Math. Gen. {\bf 39}, 12489 (2006).
	
	\bibitem{PhysRevA.96.042114} J. Liu  and H. Yuan, Control-enhanced multiparameter quantum estimation, Phys. Rev. A {\bf 96}, 042114 (2017).
	
	\bibitem{reich2012monotonically} D. M. Reich, M. Ndong, and C. P. Koch, Monotonically convergent optimization in quantum control using {K}rotov's method, J. Chem. Phys. {\bf 136}, 104103 (2012).
	
	\bibitem{PhysRevRes} D. Basilewitsch, H.  Yuan, and C. P. Koch, Optimally controlled quantum discrimination and estimation, Phys. Rev. Research {\bf 2}, 033396 (2020).
	
	\bibitem{PhysRevA.84.022326}  T. Caneva,  T.  Calarco, and S. Montangero, Chopped random-basis quantum optimization, Phys. Rev. A {\bf 84}, 022326 (2011).
	
	\bibitem{PhysRevA.102.043707} J. Tian,  H.  Liu,  Y.  Liu,  P.  Yang, R. Betzholz,  R. S.  Said,  F.  Jelezko, and J.  Cai, Quantum optimal control using phase-modulated driving fields, Phys. Rev. A {\bf 102}, 043707 (2020).
	
	
	\bibitem{xu2019generalizable} H. Xu,  J.  Li,  L.  Liu, Y.  Wang,  H.  Yuan, and X. Wang, Generalizable control for quantum parameter estimation through reinforcement learning, npj Quantum Inform. {\bf 5}, 82 (2019).
	
	
	\bibitem{PhysRevA.103.042615} H. Xu,  L.  Wang,  H. Yuan, and X. Wang, Generalizable control for multiparameter quantum metrology, Phys. Rev. A {\bf 103}, 042615 (2021).
	
	
	\bibitem{PhysRevA.68.062308} J. P. Palao and R. Kosloff, Optimal control theory for unitary transformations, Phys. Rev. A {\bf 68}, 062308 (2003).	
\end{thebibliography}
\end{document}